\documentclass[journal]{IEEEtran}

\usepackage{amsmath,amsfonts,amssymb}
\usepackage{mathtools}
\usepackage{amsthm}
\newtheorem{theorem}{Theorem}
\newtheorem{lemma}{Lemma}

\newtheorem{definition}{Definition}
\newtheorem{assumption}{Assumption}
\newtheorem{remark}{Remark}

\usepackage{algorithmic}
\usepackage{algorithm}
\usepackage{array}
\usepackage[caption=false,font=normalsize,labelfont=sf,textfont=sf]{subfig}
\usepackage{textcomp}
\usepackage{stfloats}
\usepackage{url}
\usepackage{verbatim}
\usepackage{graphicx}
\usepackage{booktabs}
\usepackage{placeins}
\usepackage{enumerate}
\usepackage{xcolor}

\usepackage[backend=bibtex,sorting=none,maxbibnames=99, minbibnames=99]{biblatex}
\usepackage{acro}

\usepackage{tikz}
\usetikzlibrary{arrows.meta,calc,positioning,fit,backgrounds}
\usepackage{pgfplots}
\usepackage{pgfplotstable}
\pgfplotsset{compat=1.18}
\usepackage[european,siunitx]{circuitikz}
\ctikzset{bipoles/thickness=1, inductors/coils=4, inductor=cute}

\definecolor{busblack}{RGB}{25,25,25}
\definecolor{gfmblue}{RGB}{0,84,159}
\definecolor{gflorange}{RGB}{204,88,3}
\definecolor{loadgray}{RGB}{95,95,95}
\definecolor{gridgreen}{RGB}{0,110,72}

\usepackage{hyperref}

\DeclareAcronym{lti}{short=LTI, long=Linear Time-Invariant}
\DeclareAcronym{ofp}{short=OFP, long=output feedback passive}
\DeclareAcronym{ifp}{short=IFP, long=input feedforward passive}
\DeclareAcronym{if-ofp}{short=IF-OFP, long=input feedforward-output feedback passive}
\DeclareAcronym{lmi}{short=LMI, long=Linear Matrix Inequality}
\DeclareAcronym{bmi}{short=BMI, long=Bilinear Matrix Inequality}
\DeclareAcronym{fdi}{short=FDI, long=Frequency-Domain Inequality}
\DeclareAcronym{kyp}{short=KYP, long=Kalman--Yakubovich--Popov}
\DeclareAcronym{gkyp}{short=gKYP, long=Generalized KYP}
\DeclareAcronym{nlsdp}{short=NL-SDP, long=Nonlinear Semidefinite Program}
\DeclareAcronym{mf}{short=MF, long=Mid-Frequency}
\DeclareAcronym{hf}{short=HF, long=High-Frequency}
\DeclareAcronym{lf}{short=LF, long=Low-Frequency}
\DeclareAcronym{vsc}{short=VSC, long=Voltage Source Converter}
\DeclareAcronym{cc}{short=CC, long=Current-Controlled}
\DeclareAcronym{pi}{short=PI, long=Proportional-Integral}
\DeclareAcronym{pll}{short=PLL, long=Phase-Locked Loop}
\DeclareAcronym{pwm}{short=PWM, long=Pulse-Width Modulation}
\DeclareAcronym{fd}{short=FD, long=Finite Difference}
\DeclareAcronym{pcc}{short=PCC, long=Point of Common Coupling}

\DeclareAcronym{sdp}{short=SDP, long=Semidefinite Program}
\newcommand{\R}{\mathbb{R}}
\renewcommand{\H}{\mathbb{H}}
\newcommand{\C}{\mathbb{C}}
\newcommand{\PhiCT}{\begin{bmatrix} 0 & 1 \\ 1 & 0 \end{bmatrix}}
\newcommand{\PsiLF}{\Psi_{\mathrm{LF}}}
\newcommand{\PsiMF}{\Psi_{\mathrm{MF}}}
\newcommand{\PsiHF}{\Psi_{\mathrm{HF}}}
\renewcommand{\L}{\mathcal{L}}
\renewcommand{\S}{\mathcal{S}}
\newcommand{\h}{\mathcal{H}}
\newcommand{\m}{\mathcal{M}}

\newcommand{\Ifp}{\nu'}
\newcommand{\Ofp}{\rho'}

\begin{document}

\title{Passivity Shaping over Frequency Bands: A Differentiable Optimization Approach}

\author{Alberto De Luca, Pol Jane-Soneira
\thanks{Alberto De Luca is with ETH Zurich, Zurich, Switzerland (e-mail: adeluc@ethz.ch). This work was performed while he was with ABB Corporate Research, Mannheim, Germany.}%
\thanks{Pol Jane-Soneira is with ABB Corporate Research, Mannheim, Germany (e-mail: pol.jane-soneira@de.abb.com).}%
\thanks{This work has been submitted to the IEEE for possible publication.
Copyright may be transferred without notice, after which this version may no
longer be accessible.}}



\maketitle

\begin{abstract}
The energy transition replaces traditional inertial synchronous machines with power electronic converters, 
introducing new grid stability challenges.
Passivity provides a compositional input--output framework for assessing stability in interconnected grid subsystems, 
yet tuning controllers for this property over specific frequency ranges---a practical requirement in converter-dominated grids---still 
relies largely on practitioners' experience, typically based on intuition or trial and error.
We address this gap in three steps: (i) we formulate the passivity index over arbitrary frequency bands 
as the optimal value of a parametric Semidefinite Program (SDP) using the generalized KYP (gKYP) lemma;
(ii) we derive analytical sensitivities of this optimal value with respect to controller parameters via the Envelope Theorem; 
and (iii) we embed these elements in a bilevel optimization 
scheme that tunes controller parameters to maximize the passivity index.
We validate the proposed method on two case studies of increasing complexity:
a grid-following converter and a multi-converter microgrid. 
\end{abstract}

\section{Introduction}
\IEEEPARstart{P}{ower} systems are undergoing a fundamental transition from synchronous-machine-dominated grids to 
converter-based grids, 
driven by the large-scale integration of renewable energy sources~\cite{Bahrani2024}. 
This shift introduces new challenges in stability analysis. 
The mechanical inertia of synchronous generators, which has traditionally contributed to moderating the rate of change of system frequency and 
shaping electromechanical oscillatory behavior, is progressively diminishing. 
Consequently, power systems become more sensitive to fast dynamics, and higher-frequency interactions that were previously less significant may gain importance~\cite{Dorfler2023,Kong2022}. 
Dissipativity~\cite{vanderschaft2017dissipative}, and more specifically passivity~\cite{Willems1972a,Willems1972b}, is widely regarded as a promising tool 
for the stability analysis of such power systems. 
The key advantage is that it offers a compositional framework to guarantee stability under interconnection at the subsystem level. 
Thus, the need for deriving and analyzing a model of the whole system, which is impractical or even impossible due to privacy considerations, is alleviated.

The practical relevance of passivity-based requirements is starting to be recognized at the regulatory level. 
In August 2024, the non-passive behavior of an offshore converter connection triggered a harmonic instability 
event in the German transmission system~\cite{BNetzA2025}. 
The German regulator consequently recommended introducing passivity over the relevant frequency range as a mandatory grid-connection 
requirement for all new assets~\cite[p.~53]{BNetzA2025}.
This trend extends beyond Germany: the European Network of Transmission System 
Operators for Electricity (ENTSO-E) recommends in its technical guidance on grid-forming 
requirements that power generating units exhibit passive impedance behavior over a 
frequency range specified by the transmission system 
operator~\cite[p.~21]{ENTSOE_TGGFC2025}. Likewise, NERC recommends that grid-forming inverters should be passive at $0 - 300$ Hz \cite{nerc2023gfm_bess}, and UNIFI recommends a combination of passivity and low gain in that frequency range~\cite{kroposki2025unifi}. Fingrid, Finland's transmission system operator, has already updated its grid code to require grid-forming battery energy storage systems to exhibit passive behavior in the frequency ranges of 0-47 Hz and 53-250 Hz as a condition for grid connection~\cite{haapaniemi2023fingrid_bess}.

Despite this regulatory push, passivity-oriented controller tuning in power-system applications remains challenging in practice.
In many practical designs, controller parameters are still selected through engineering intuition or parameter sweeps~\cite{RodriguezDiaz2019},
even though passivity violations at specific frequencies 
often lead to instability in grid-connected converters~\cite{Harnefors2016Passivity}. The difficulty is inherent to \acp{vsc}, which exhibit negative-conductance behavior 
in certain frequency ranges, making full-spectrum passivity unattainable. Moreover, passivity margins depend on controller
gains in non-monotonic and counterintuitive ways~\cite{Chen2024}, which makes parameter sweeps unreliable and manual tuning 
impractical, especially when many parameters must be adjusted at the same time.

Several prior works address passivity-oriented controller design, each focusing on a specific problem setting. 
Feedback passivation has been formulated as a convex optimization problem 
with theoretical guarantees~\cite{Su2020}; however, this approach considers
only the standard feedback interconnection and is not applicable to more complex control structures
composed of inner and outer loops, as is typical in power system applications. Furthermore, it targets passivity over the entire frequency axis, not over specific frequency bands.
The passivity sensitivity methodology based on admittance models presented in~\cite{Lee2025} computes how individual parameters influence the passivity index at each frequency.
However, this approach provides only pointwise sensitivities and does not capture the worst-case passivity margin over a frequency band.
Recently, the authors of~\cite{kowalewski2025passivity} derived two frequency-dependent passivity-based stability 
criteria for the converter--grid interconnection, further motivating passivity shaping over
targeted frequency bands rather than over the entire frequency spectrum.
Moreover, the work in~\cite{peng2026positive} formulates finite-frequency passivity
requirements as \ac{lmi} constraints via the \ac{gkyp} lemma and provides a 
visual tool for analyzing the feasibility of passivization. We build on these
advances: we formalize the general problem of tuning controllers to maximize
passivity over targeted frequency ranges as an optimization problem, exploit
its structure, and derive a gradient-based algorithm applicable to any
differentiable controller parameterization.
The proposed optimization-based formulation results in a non-convex problem: while the passivity index computation for fixed parameters constitutes a convex problem, 
jointly optimizing over the controller parameter space introduces non-convexity.
The passivity index is not an explicit function of the controller parameters; rather, it is the optimal value of a parametric \ac{sdp}. 
Consequently, passivity-oriented tuning requires computing how the optimal value of the \ac{sdp} changes when the controller parameters vary.

Differentiable optimization~\cite{bonnans2000perturbation} provides a natural framework for this task, offering tools to compute how optimal values change with problem parameters.
Recent work has used these ideas to embed optimization problems inside larger learning and control architectures. 
For example, quadratic programs have been implemented as differentiable layers in neural networks~\cite{Amos2017}, and this idea has been extended to disciplined convex programs together with open-source software support~\cite{Agrawal2019}. 
In control, related techniques have been used to differentiate through predictive-control policies, enabling controller parameters to be tuned through the policy's underlying optimization problem~\cite{Zuliani2025}. 
We build upon these advances to address a
different objective: 
tuning controller parameters to improve passivity margins for \ac{lti} systems.

Consequently, the contribution of this work is threefold:
\begin{enumerate}
   \item We formulate controller tuning as the maximization of the
    passivity index over a prescribed frequency band through the \ac{gkyp}
    lemma. We show that, for fixed controller parameters, this index is the optimal
    value of an \ac{sdp} whose data depend on those parameters.

    \item We derive analytical sensitivities of this optimal value with
    respect to the controller parameters by differentiating through the
    \ac{sdp}.

    \item We combine the above into a bilevel algorithm that applies to
    arbitrary differentiable controller parameterizations and preserves
    closed-loop stability at every iterate.
\end{enumerate}

The remainder of this paper is organized as follows. 
Section~\ref{sec:prob_form} introduces the required preliminaries. 
Section~\ref{sec:envelope} formulates the passivity-oriented tuning problem and derives the analytical sensitivities of passivity indices over prescribed frequency ranges. 
Section~\ref{sec:bilevel} develops the proposed optimization algorithm. 
Section~\ref{sec:simulations} validates the framework on two case studies of increasing complexity.
Section~\ref{sec:conclusion} concludes the paper and outlines directions for future work.

\emph{Notation:}
We denote by $\mathbb{R}$ and $\mathbb{C}$ the sets of real and complex numbers, respectively. 
The sets of real symmetric and complex Hermitian $n\times n$ matrices are denoted by $\mathbb{S}^n$ and $\mathbb{H}^n$,
respectively. For vectors or matrices $X,Y$, $\langle X,Y\rangle$ denotes the inner product and $X^\top$ the transpose; for matrices,
$\langle X,Y\rangle := \operatorname{tr}(X^*Y)$, where $\operatorname{tr}(\cdot)$ is the trace operator. 
For a matrix $X$, $X^*$ denotes its Hermitian adjoint, and $X^{-1}$ its inverse, when it exists.
We also write $X^{-*}:=(X^{-1})^*$. We write $A\succeq0$ ($A\succ0$) to indicate that a matrix is positive semidefinite (positive definite),
and $A\preceq0$ ($A\prec0$) for negative semidefinite (negative definite). For a square matrix $X$, $\lambda_{\min}(X)$ denotes its smallest eigenvalue.
$I_n$ denotes the $n\times n$ identity matrix.
The superscript $^\star$ denotes optimality: for an optimization problem, $x^\star$ denotes an optimizer and $f^\star := f(x^\star)$ the corresponding optimal value.
\section{Preliminaries}\label{sec:prob_form}

\subsection{Parametric LTI Model}
We consider a parametric \ac{lti} dynamical system whose 
state-space realization depends on a vector of tunable parameters 
$\theta \in \Theta \subseteq \R^p$:
\begin{equation}\label{eq:parameterized_cl}
\dot{x} = A(\theta)\,x + B(\theta)\,u, \qquad
y = C(\theta)\,x + D(\theta)\,u,
\end{equation}
with state $x \in \R^n$, input $u \in \R^m$, output 
$y \in \R^m$, and parameter-dependent state-space matrices 
$A \colon \Theta \to \R^{n \times n}$, 
$B \colon \Theta \to \R^{n \times m}$, 
$C \colon \Theta \to \R^{m \times n}$, 
$D \colon \Theta \to \R^{m \times m}$.
Note that system~\eqref{eq:parameterized_cl} may arise from the linearization of a nonlinear system. The next assumption restricts how the parameters may influence the system matrices.
\begin{assumption}
    The entries of $A, B, C, D$ are continuously differentiable on the tunable parameters $\theta$.
\end{assumption}
Note that no assumption is made about how exactly the tunable parameters enter into the system matrices. We allow hence, in contrast to the literature, to tune parameters of controllers in cascaded or nonlinear structures. The entries of the system matrices may be nonlinear functions of $\theta$, the only requirement is that they are continuously differentiable, which is the case when there are no discontinuities in the system or controllers.
The corresponding input--output transfer function is
\begin{equation}
\label{eq:transfer}
G_{\theta}(s) := C(\theta)\bigl(sI - A(\theta)\bigr)^{-1}B(\theta) + D(\theta),
\end{equation}
where $s \in \C$ denotes the Laplace variable.
In power system applications, \(\theta\) may be, for example, 
the parameters of the \ac{pi} controllers of the asset to be tuned,
while \(u\) and \(y\) denote current and voltage for interconnection with the grid, depending on the choice of impedance or admittance model.
We note, however, that the presented approach is not limited to power system applications.
The goal of this work is to tune the parameters $\theta \in \Theta$ such that the passivity properties of system~\eqref{eq:parameterized_cl} (or, equivalently,~\eqref{eq:transfer}) 
are optimized. In the following two subsections, we present the basics of passivity theory in the time and frequency domain. 

\subsection{Time-Domain Passivity}
We recall the definition of dissipativity for system~\eqref{eq:parameterized_cl} 
and extend it to passivity.

\begin{definition}[Dissipativity~\cite{Hill1976}]
Given a supply rate \(w \colon \mathbb{R}^{m} \times \mathbb{R}^{m} \to \mathbb{R}\),
System~\eqref{eq:parameterized_cl} is \emph{dissipative} with respect to \(w\) if
\[
  \int_{0}^{t} w\bigl(u(\tau), y(\tau)\bigr) \,\mathrm{d}\tau \geq 0
\]
for all \(t \geq 0\) and all admissible inputs \(u(\cdot)\),
where \(y(\cdot)\) is the output of~\eqref{eq:parameterized_cl} along the trajectory
generated by \(u(\cdot)\) from \(x(0) = 0\).
\end{definition}


\begin{definition}[Passivity and variants] \label{def:passivity}
    Consider the supply rate
    \begin{equation*}
    w_{\textnormal{P}}(u, y) \coloneqq -\begin{bmatrix} y \\ u \end{bmatrix}^\top \mathcal{S}(\rho', \nu') \begin{bmatrix} y \\ u \end{bmatrix},
    \end{equation*}
    where
    \begin{equation}\label{eq:supply_rate_passivity}
    \S(\rho',\nu') :=
    \begin{bmatrix}
    \rho'\, I_m & -\tfrac{1}{2}\,I_m \\[2pt]
    -\tfrac{1}{2}\,I_m & \nu'\, I_m
    \end{bmatrix} \in \mathbb{S}^{2m},
    \end{equation}
    with constants $\Ifp, \Ofp \in \R$. System \eqref{eq:parameterized_cl} is called
    \begin{itemize}
        \item[(i)] passive if it is dissipative with respect to the supply rate $w_{\textnormal{P}}(u,y)$ with $\Ifp \geq 0$ and $\Ofp \geq 0$.
        \item[(ii)] \ac{ifp} with index $\Ifp\in \R$ if it is dissipative with respect to the supply rate $w_{\textnormal{P}}(u,y)$ with $\Ofp = 0$.
        \item[(iii)] \ac{ofp} with index $\Ofp \in \R$ if it is dissipative with respect to the supply rate $w_{\textnormal{P}}(u,y)$ with $\Ifp = 0$.
        \item[(iv)] \ac{if-ofp} if it is dissipative with respect to the supply rate $w_{\textnormal{P}}(u,y)$ with $\Ifp \in \R$ and $\Ofp \in \R$. 
    \end{itemize}
\end{definition}


\subsection{Frequency-Domain Passivity}

Passivity admits an equivalent frequency-domain characterization 
as a \ac{fdi} on the transfer function~\eqref{eq:transfer} for stable \ac{lti} systems~\cite{Bao2007}.

\begin{definition}[Positive realness]
\label{def:positive_real}
The transfer function $G_{\theta}(s)$ in~\eqref{eq:transfer} is 
\emph{$(\nu,\rho)$-positive real} on a set $\Omega \subseteq \mathbb{R}$, 
with scalars $\nu, \rho \in \mathbb{R}$, if
\begin{equation}\label{eq:fdi_passivity}
\begin{bmatrix} G_{\theta}(j\omega) \\ I_m\end{bmatrix}^{\!*}
\S(\rho,\nu)
\begin{bmatrix} G_{\theta}(j\omega) \\ I_m \end{bmatrix}
\preceq 0,
\quad \forall\,\omega \in \Omega,
\end{equation}
with \(\S(\rho,\nu)\) as in~\eqref{eq:supply_rate_passivity}.

\end{definition}

When $\Omega = \mathbb{R}$, the well-known \ac{kyp} lemma~\cite{rantzer1996kyp} establishes 
that $(\nu,\rho)$-positive realness of $G_{\theta}(s)$ is equivalent to 
\ac{if-ofp} of system~\eqref{eq:parameterized_cl} with 
$\nu = \nu'$ and $\rho = \rho'$~\cite[Remark~4.1.6]{vanderschaft2017dissipative}.

Condition~\eqref{eq:fdi_passivity} is a \ac{fdi} that must hold at every $\omega \in \Omega$. The \ac{gkyp} Lemma~\cite{iwasaki2005generalized} converts this semi-infinite constraint into a single finite-dimensional \ac{lmi} for arbitrary frequency intervals $\Omega \subseteq \mathbb{R}$, which is introduced next.

\subsection{Generalized KYP Lemma}

Since the matrices in~\eqref{eq:parameterized_cl} are restricted to real entries,
we adopt the continuous-time specialization of the \ac{gkyp} lemma presented in~\cite{pipeleers2011generalized}.

\begin{theorem}[\ac{gkyp} Lemma with Real Data~{\cite{pipeleers2011generalized}}]
\label{thm:gkyp_real}
Let $A \in \R^{n \times n}$, $B \in \R^{n \times m}$, and 
$\Sigma \in \S^{n+m}$. Suppose $(A,B)$ is controllable and 
$\det(j\omega I_n - A) \neq 0$ for all $\omega \in \Omega \subseteq \R$. 
Let $\Psi_{\Omega} \in \H^{2}$ be the Hermitian matrix associated with 
the frequency set $\Omega$. Then the following statements are equivalent:
\begin{enumerate}[(i)]
    \item The \ac{fdi} holds for all $\omega \in \Omega$:
    \begin{equation}\label{eq:fdi_gkyp}
        \begin{bmatrix} (j\omega I_n - A)^{-1}B \\ I_m \end{bmatrix}^{*}
        \Sigma
        \begin{bmatrix} (j\omega I_n - A)^{-1}B \\ I_m \end{bmatrix}
        \preceq 0.
    \end{equation}
    \item There exist $P,\, Q \in \mathbb{S}^{n}$ with $Q \succeq 0$ 
    such that
    \begin{equation}\label{lmi:gkyp_rd}
        \begin{bmatrix} A & B \\ I_n & 0 \end{bmatrix}^{\!\top}
        \!\left(\PhiCT \otimes P + \Psi_{\Omega} \otimes Q\right)
        \begin{bmatrix} A & B \\ I_n & 0 \end{bmatrix}
        + \Sigma \preceq 0.
    \end{equation}
\end{enumerate}
\end{theorem}

The matrix $\Psi_{\Omega}$ encodes the frequency range of interest. 
Table~\ref{tab:psi_cases} summarizes the \ac{lf}, \ac{mf}, 
and \ac{hf} cases for continuous-time 
systems~\cite{iwasaki2005generalized}.

\begin{table}[!h]
\centering
\caption{Frequency ranges and associated $\Psi_{\Omega}$ 
matrices~\cite{iwasaki2005generalized}.}
\label{tab:psi_cases}
\begin{tabular}{lcc}
\toprule
Case & Frequency set $\Omega$ & $\Psi_{\Omega}$ \\
\midrule
LF & $|\omega| \leq \bar{\omega}_{\ell}$ & 
$\begin{bmatrix} -1 & 0 \\ 0 & \bar{\omega}_{\ell}^2 \end{bmatrix}$ \\[6pt]
MF & $\bar{\omega}_1 \leq \omega \leq \bar{\omega}_2$ & 
$\begin{bmatrix} -1 & j\bar{\omega}_c \\ -j\bar{\omega}_c & -\bar{\omega}_1 \bar{\omega}_2 \end{bmatrix}$ \\[6pt]
HF & $|\omega| \geq \bar{\omega}_h$ & 
$\begin{bmatrix} 1 & 0 \\ 0 & -\bar{\omega}_h^2 \end{bmatrix}$ \\
\bottomrule
\end{tabular}

\smallskip
$\bar{\omega}_c = (\bar{\omega}_1 + \bar{\omega}_2)/2$
\end{table}

Substituting the parametric realization~\eqref{eq:parameterized_cl} and the passivity supply rate $\S(\rho,\nu)$ into~\eqref{lmi:gkyp_rd} yields
\begin{equation}\label{eq:M_omega}
\begin{split}
M_{\Omega}(P,Q,\rho,\nu;\theta) &:=
\begin{bmatrix} A(\theta) & B(\theta) \\ I_n & 0 \end{bmatrix}^{\!\top}
\!\left(\PhiCT \otimes P \right. \\
&\left.+ \Psi_{\Omega} \otimes Q\right)
\begin{bmatrix} A(\theta) & B(\theta) \\ I_n & 0 \end{bmatrix}
+ \Sigma(\rho,\nu;\theta) \preceq 0,
\end{split}
\end{equation}
where the semicolon separates the \ac{lmi} variables from the parameter vector~$\theta$, and
\begin{equation}\label{eq:sigma_passivity}
\Sigma(\rho,\nu;\theta) = 
\begin{bmatrix}
    C(\theta) & D(\theta) \\
    0         & I_m
\end{bmatrix}^{\!\top}
\S(\rho, \nu)
\begin{bmatrix}
    C(\theta) & D(\theta) \\
    0         & I_m
\end{bmatrix}.
\end{equation}

\begin{remark}
    \label{rem:gkyp_ifp_ofp}
    The \ac{gkyp} Lemma is a general result with applications beyond passivity; with the supply rate~\eqref{eq:supply_rate_passivity}, it characterizes the \ac{ifp} and \ac{ofp} indices of~$G_\theta$ over a frequency range~$\Omega$.
    The largest~$\nu$ that satisfies~\eqref{eq:fdi_passivity} with $\rho = 0$ defines the \ac{ifp} index~\cite{Bao2007}
    \begin{equation*}
    \nu^{\star} = \inf_{\omega\in\Omega}\, \lambda_{\min}\!\left(
    \tfrac{1}{2}\bigl(G_{\theta}(j\omega)+G_{\theta}(j\omega)^{*}\bigr)
    \right).
    \end{equation*}
    Likewise, the largest~$\rho$ that satisfies~\eqref{eq:fdi_passivity} with $\nu = 0$ defines the \ac{ofp} index
    \begin{equation*}
    \rho^{\star} = \inf_{\omega\in\Omega}\, \lambda_{\min}\!\left(
    \tfrac{1}{2}\bigl(G_{\theta}(j\omega)^{-1}+G_{\theta}(j\omega)^{-*}\bigr)
    \right),
    \end{equation*}
    provided that $G_{\theta}(j\omega)$ is nonsingular for all $\omega\in\Omega$. Theorem~\ref{thm:gkyp_real} replaces \eqref{eq:fdi_passivity} with the equivalent \ac{lmi}~\eqref{lmi:gkyp_rd}; maximizing~$\nu$ or~$\rho$ subject to this \ac{lmi} therefore yields these indices.
\end{remark}

A positive $\nu^{\star}$ signals excess passivity at the input port; a positive $\rho^{\star}$ at the output port. Negative values indicate a shortage at the respective port. These indices go beyond a binary passive/non-passive classification: they quantify how much passivity each port lacks or exceeds over a specific frequency range, taking the worst-case value as the measure~\cite{Bao2007}.

\subsection{Envelope Theorem}\label{subsec:envelope_prelim}
Differentiable optimization studies how optimizers and the corresponding optimal value respond to perturbations of the problem data. For parametric \ac{sdp}s, the Envelope Theorem~\cite{milgrom2002envelope} expresses that response in closed form for the optimal value, in terms of an optimal primal--dual pair. Consider a parametric \ac{sdp}:
\begin{equation}
\label{opt:parametric_sdp}
\begin{aligned}
v^{\star}(\theta) \;:=\; \min_{x}\quad & c(\theta)^{\!\top} x \\[4pt]
\text{s.t.}\quad 
& F_0(\theta) + \sum_{i=1}^{d} x_i\, F_i(\theta) \succeq 0,
\end{aligned}
\end{equation}
with optimization variables $x \in \R^d$, parameters $\theta \in \Theta \subseteq \R^p$, $c \colon \Theta \to \R^d$, and $F_i \colon \Theta \to \mathbb{S}^{q}$, $i = 0,\ldots,d$. 
The Lagrangian\footnote{See~\cite{Shapiro1997} for Lagrangian duality in \ac{sdp}.} is
\begin{equation*}
L(x, Z;\, \theta) := c(\theta)^{\!\top} x 
- \left\langle Z,\; F_0(\theta) + \sum_{i=1}^{d} x_i\, F_i(\theta)
\right\rangle,
\end{equation*}
with dual multiplier $Z \in \S^{q}$, $Z \succeq 0$.

The next assumption is a technical requirement for the Envelope Theorem presented below.

\begin{assumption}[Regularity]
\label{ass:local_regularity}
In a neighborhood of~$\bar{\theta} \in \Theta$:
(i)~$c(\theta)$ and $F_i(\theta)$, $i=0,\ldots,d$, are continuously
differentiable;
(ii)~Slater's condition holds at~$\bar{\theta}$, i.e., there exists
$\tilde{x} \in \mathbb{R}^d$ with
$F_0(\bar{\theta}) + \sum_{i=1}^{d} \tilde{x}_i F_i(\bar{\theta}) \succ 0$,
and the optimal value $v^{\star}(\theta)$ is finite and attained for all
$\theta$ in this neighborhood;
(iii)~let $\mathcal{X}^{\star}(\bar\theta)$ and $\Lambda^{\star}(\bar\theta)$
denote the primal and dual optimal sets of~\eqref{opt:parametric_sdp}. For each
$k = 1,\ldots,p$, the quantity
\begin{equation*}
\frac{\partial c(\bar\theta)^{\!\top}}{\partial \theta_k} x
- \left\langle Z,\; \frac{\partial F_0(\bar\theta)}{\partial \theta_k}
+ \sum_{i=1}^{d} x_i\, \frac{\partial F_i(\bar\theta)}{\partial \theta_k} \right\rangle
\end{equation*}
takes the same value for all
$(x,Z) \in \mathcal{X}^{\star}(\bar{\theta}) \times \Lambda^{\star}(\bar{\theta})$.
\end{assumption}

Parts~(i) and~(ii) are standard in the sensitivity analysis of convex programs~\cite{bonnans2000perturbation}. Part~(iii) means that in case the primal-dual optimum is not unique, different optimal primal--dual pairs must predict the same first-order change in the optimal value. This assumes, in essence, differentiability of the value function with respect to the tunable parameters. We further comment on this in Remarks~\ref{rem:directional_derivative} and \ref{rem:nonsmooth}.

\begin{theorem}[Envelope Theorem~{\cite{milgrom2002envelope,bonnans2000perturbation}}]
\label{thm:envelope_sdp}
Let Assumption~\ref{ass:local_regularity} hold. Then $v^{\star}$ is differentiable
at~$\bar\theta$ and, for \emph{any}
$(x^{\star}, Z^{\star}) \in \mathcal{X}^{\star}(\bar\theta) \times \Lambda^{\star}(\bar\theta)$,
\begin{equation}\label{eq:envelope_gradient}
\begin{aligned}
\nabla_\theta v^{\star}(\bar{\theta})
&= \nabla_\theta c(\bar{\theta})^{\!\top} x^{\star} \\
&\quad- \left\langle Z^{\star},\;
\nabla_\theta F_0(\bar{\theta}) + \sum_{i=1}^{d}
x_i^{\star}\, \nabla_\theta F_i(\bar{\theta})
\right\rangle.
\end{aligned}
\end{equation}
\end{theorem}

Expression~\eqref{eq:envelope_gradient} provides the sensitivity of the 
optimal value with respect to $\theta$ in closed form using only an optimal primal--dual 
solution and the partial derivatives of the problem data with respect to $\theta$.
It avoids both numerical differentiation and implicit differentiation through 
the solver's optimality conditions.

\begin{remark}
\label{rem:directional_derivative}
    When part (iii) of Assumption~\ref{ass:local_regularity} is not satisfied, the optimal value function may fail to be differentiable in the classical sense. Directional differentiability offers a framework for sensitivity analysis also in this case, see~\cite{Shapiro1997}; a detailed treatment of the nonsmooth case is beyond the scope of this work.
\end{remark}

At first glance, this general result from differentiable optimization may appear unrelated to passivity. In Section~\ref{sec:envelope}, we show how to exploit it as the key computational tool for passivity-oriented controller tuning.
\section{Passivity Index Optimization and Sensitivities}\label{sec:envelope}
This section formulates the passivity-index optimization problem and derives the sensitivities of its optimal value with respect to the tunable controller parameters. We first formulate the frequency-domain optimization problem and derive its \ac{gkyp}-based matrix inequality representation. We then specialize this representation to the frequency ranges of interest and apply the Envelope Theorem to obtain closed-form sensitivity expressions.
\subsection{Passivity Index Optimization}
The goal of passivity-index optimization is to determine a parameter vector $\theta \in \Theta$ that optimizes the passivity properties in~\eqref{eq:fdi_passivity} of system~\eqref{eq:parameterized_cl} over a frequency interval~$\Omega$. The resulting optimization problem takes the form
\begin{equation}
\label{opt:non_finite_passivity}
\begin{aligned}
\max_{\theta,\,\rho,\,\nu}\qquad & \hspace{6em} \alpha(\rho,\nu) \\
\text{s.t.}\quad & \hspace{1em} \theta \in \Theta, \\
& \begin{bmatrix} G_{\theta}(j\omega) \\ I_m \end{bmatrix}^{\!*}
\S(\rho,\nu)
\begin{bmatrix} G_{\theta}(j\omega) \\ I_m \end{bmatrix}
\preceq 0,
\quad \forall\,\omega \in \Omega,
\end{aligned}
\end{equation}
with $\S(\rho,\nu)$ as in~\eqref{eq:supply_rate_passivity} and
\begin{equation}\label{eq:alpha_def}
\alpha(\rho,\nu) :=
\begin{cases}
\rho, & \text{(\acs{ofp} case, $\nu = \bar{\nu} = 0$),} \\
\nu, & \text{(\acs{ifp} case, $\rho =\bar{\rho} = 0$).}
\end{cases}
\end{equation}
Problem~\eqref{opt:non_finite_passivity} is computationally 
intractable: the constraint must hold for every frequency \(\omega\) in a continuum \(\Omega\) (e.g., \(\R\) or
\([\omega_1, \omega_2]\)).
 Applying the \ac{gkyp} lemma (Theorem~\ref{thm:gkyp_real}) replaces the semi-infinite constraint with a single finite-dimensional \ac{lmi}:
\begin{equation}
\label{opt:non_convex_passivity}
\begin{aligned}
\max_{\theta,\,P,\,Q,\,\rho,\,\nu}\qquad & \alpha(\rho,\nu)\\
\text{s.t.}\quad & \theta \in \Theta,\quad Q \succeq 0, \\
& M_{\Omega}(P,Q,\rho,\nu,\theta) \preceq 0,
\end{aligned}
\end{equation}
where $M_{\Omega}$ denotes the \ac{lmi} in~\eqref{lmi:gkyp_rd} as formulated in~\eqref{eq:M_omega}. Problem~\eqref{opt:non_convex_passivity} is non-convex: the parameter-dependent state-space matrices $A(\theta)$, $B(\theta)$, $C(\theta)$, $D(\theta)$ enter $M_{\Omega}$ as optimization variables through~$\theta$ and multiply the decision variables $P$ and $Q$, making the constraint~\eqref{eq:M_omega} a nonlinear matrix inequality in the joint variables $(\theta, P, Q)$. For fixed~$\theta$, the problem reduces to a standard \ac{sdp}.

\subsection{Specialization to Frequency Ranges}
\label{sec:freq_ranges}

In the following, we specialize $M_{\Omega}$ in \ac{lmi}~\eqref{eq:M_omega} for each frequency range 
in Table~\ref{tab:psi_cases}.

For ease of notation, we drop the explicit dependence on $\theta$ 
from the state-space matrices $A, B, C, D$.
In each case, the \ac{lmi} takes the form~\eqref{eq:M_omega}
with $\Sigma$ as defined in~\eqref{eq:sigma_passivity} partitioned as
\begin{equation*}
\Sigma = 
\begin{bmatrix}
\Sigma_{11} & \Sigma_{12} \\
\Sigma_{21} & \Sigma_{22}
\end{bmatrix},
\end{equation*}
where, expanding the matrix products,
\begin{align*}
\Sigma_{11} &= \rho\, C^\top C, \\
\Sigma_{12} &= -\tfrac{1}{2} C^\top + \rho\, C^\top D, \\
\Sigma_{22} &= -\tfrac{1}{2}(D + D^\top) + \rho\, D^\top D + \nu\, I_m,
\end{align*}
and $\Sigma_{21} = \Sigma_{12}^\top$.
We select $\Psi_{\Omega}$ from Table~\ref{tab:psi_cases} according to 
the frequency range of interest in the following cases.

\subsubsection{Low-Frequency Case}

For the \ac{lf} range ($|\omega| \leq \bar{\omega}_\ell$), the middle matrix in the first term of \ac{lmi}~\eqref{eq:M_omega} specializes to
\begin{equation*}
\PhiCT \otimes P + \PsiLF \otimes Q
=
\begin{bmatrix} -Q & P \\ P & \bar{\omega}_\ell^{\,2}\, Q \end{bmatrix}.
\end{equation*}
Expanding the matrix products yields the block entries of the first term of~\eqref{eq:M_omega}:
\begin{align*}
\L_{11} &= A^{\!\top}\!P + PA - A^{\!\top}\!QA 
+ \bar{\omega}_\ell^{\,2}\, Q, \\
\L_{12} &= PB - A^{\!\top}\!QB, \\
\L_{22} &= -B^{\!\top}\!QB,
\end{align*}
with \(\L_{21} = \L_{12}^{\!\top}\). Adding the passivity supply-rate matrix $\Sigma$, the \ac{gkyp} \ac{lmi} \eqref{eq:M_omega} becomes
\begin{equation*}
M_{\text{LF}}:=
\begin{bmatrix}
\L_{11} + \Sigma_{11} & \L_{12} + \Sigma_{12} \\
\L_{21} + \Sigma_{21} & \L_{22} + \Sigma_{22}
\end{bmatrix}
\preceq 0.
\end{equation*}

\subsubsection{High-Frequency Case}

For the \ac{hf} range ($|\omega| \geq \bar{\omega}_h$), the middle matrix in the first term of \ac{lmi}~\eqref{eq:M_omega} specializes to
\begin{equation*}
\PhiCT \otimes P + \PsiHF \otimes Q
=
\begin{bmatrix} Q & P \\ P & -\bar{\omega}_h^{\,2}\, Q \end{bmatrix}.
\end{equation*}
Expanding the matrix products yields the block entries of the first term of~\eqref{eq:M_omega}:
\begin{align*}
\h_{11} &= A^{\!\top}\!P + PA + A^{\!\top}\!QA 
- \bar{\omega}_h^{\,2}\, Q,\\
\h_{12} &= PB + A^{\!\top}\!QB,\\
\h_{22} &= B^{\!\top}\!QB,
\end{align*}
with \(\h_{21} = \h_{12}^{\!\top}\). Adding the passivity supply-rate matrix $\Sigma$ produces the high-frequency condition
\begin{equation*}
M_{\text{HF}}:=
\begin{bmatrix}
\h_{11} + \Sigma_{11}
& \h_{12} + \Sigma_{12} \\
\h_{21} + \Sigma_{21}
& \h_{22} + \Sigma_{22}
\end{bmatrix}
\preceq 0.
\end{equation*}

\subsubsection{Mid-Frequency Case}

For the \ac{mf} range ($\bar{\omega}_1 \leq \omega \leq \bar{\omega}_2$), with $\bar{\omega}_c = (\bar{\omega}_1 + \bar{\omega}_2)/2$, the middle matrix in the first term of \ac{lmi}~\eqref{eq:M_omega} specializes to
\begin{equation*}
\PhiCT \otimes P + \PsiMF \otimes Q
=
\begin{bmatrix} -Q & P + j\bar{\omega}_c\, Q \\ 
P - j\bar{\omega}_c\, Q & -\bar{\omega}_1\bar{\omega}_2\, Q \end{bmatrix}.
\end{equation*}
Expanding the matrix products yields the block entries of the first term of~\eqref{eq:M_omega}:
\begin{align*}
\m_{11} &= A^{\!\top}\!P + PA - A^{\!\top}\!QA 
+ j\bar{\omega}_c(A^{\!\top}\!Q - QA) 
- \bar{\omega}_1\bar{\omega}_2\, Q, \\
\m_{12} &= PB - A^{\!\top}\!QB - j\bar{\omega}_c\, QB, \\
\m_{22} &= -B^{\!\top}\!QB,
\end{align*}
with \(\m_{21} = \m_{12}^{*}\). Adding the passivity supply-rate matrix $\Sigma$ produces the mid-frequency condition
\begin{equation*}
M_{\text{MF}}:=
\begin{bmatrix}
\m_{11} + \Sigma_{11}
& \m_{12} + \Sigma_{12} \\
\m_{21} + \Sigma_{21}
& \m_{22} + \Sigma_{22}
\end{bmatrix}
\preceq 0.
\end{equation*}

The \ac{lf} and \ac{hf} cases yield real symmetric matrices, whereas the \ac{mf} case introduces imaginary off-diagonal coefficients that make $M_{\text{MF}}$ Hermitian. Since standard real-valued solvers cannot handle Hermitian \acp{lmi} directly, the \ac{mf} problem requires a real-valued reformulation that doubles the matrix dimension 
(Lemma~\ref{lem:complex_to_real} in Appendix). We derive the passivity-index sensitivities for the \ac{lf} and \ac{hf} cases in the next sections and refer for the \ac{mf} derivation to subsection~\ref{app:mf_sensitivity} of the Appendix.

\subsection{Sensitivity of the Passivity Index}\label{subsec:envelope_gkyp}

For a fixed $\bar{\theta} \in \Theta$, 
problem~\eqref{opt:non_convex_passivity} reduces to a standard \ac{sdp}:
\begin{equation}\label{eq:inner_sdp}
    \begin{aligned}
        \alpha^{\star}(\bar{\theta})
        = \max_{P,Q,\rho,\nu}\quad & \alpha(\rho,\nu) \\
        \text{s.t.}\quad &
        Q \succeq 0, \quad
        -M_{\Omega}(P,Q,\rho,\nu,\bar{\theta}) \succeq 0,
    \end{aligned}
\end{equation}
where $\Omega$ selects the frequency range (\ac{lf}, \ac{mf}, \ac{hf}) and 
$\alpha$ is defined in~\eqref{eq:alpha_def}.

Now, the following question arises: how does the optimal passivity index $\alpha^\star$ change when the tunable parameters $\theta$ vary? Observe that problem~\eqref{eq:inner_sdp} matches the parametric \ac{sdp} structure of~\eqref{opt:parametric_sdp}: $\theta$ acts as the external parameter, entering only through the data matrices 
$A(\theta)$, $B(\theta)$, $C(\theta)$, $D(\theta)$ that compose $M_\Omega$, while the decision variables $(P, Q, \rho, \nu)$ 
correspond to $x$ in~\eqref{opt:parametric_sdp} and are optimized for each fixed~$\theta$. Theorem~\ref{thm:envelope_sdp} can be therefore readily applied.

Let $Z_M \succeq 0$ denote the dual multiplier associated with the constraint $-M_{\Omega} \succeq 0$.
Since only $M_{\Omega}$ depends explicitly on~$\theta$, applying Theorem~\ref{thm:envelope_sdp} to the equivalent 
minimization of $-\alpha(\rho,\nu)$ yields
\begin{equation}\label{eq:envelope_gkyp} 
    \frac{\partial \alpha^{\star}}{\partial \theta_k}\bigg|_{\bar{\theta}}=
    -\left\langle
    Z_M^{\star}(\bar{\theta}),\;
    \frac{\partial M_{\Omega}}{\partial \theta_k}
    \bigg|_{\bar{\theta}}
    \right\rangle,
    \quad k=1,\dots,p.
\end{equation}

Expression~\eqref{eq:envelope_gkyp} requires only~$Z_M^{\star}(\bar{\theta})$ 
and the partial derivatives 
$\partial M_{\Omega}/\partial \theta_k|_{\bar{\theta}}$, which we 
derive next.
We adopt the shorthand 
$\partial_k A := \partial A(\theta)/\partial \theta_k|_{\bar\theta}$; 
all unsubscripted matrices $A$, $B$, $C$, $D$ and optimal 
variables $P^{\star}$, $Q^{\star}$, $\rho^{\star}$ are evaluated 
at~$\bar\theta$. Since the \ac{lf} and \ac{hf} \acp{lmi} share 
the same structure up to a sign, we unify them through
\[
\sigma := \begin{cases}
	+1, & \text{LF case}, \\
	-1, & \text{HF case}.
\end{cases}
\]

\subsubsection{Unified LF/HF Sensitivity}
Let $\Pi_k$ denote the derivative of $M_{\Omega}$ with respect to $\theta_k$, evaluated at $\bar\theta$:
\[
\Pi_k := \frac{\partial M_{\Omega}}{\partial \theta_k}\bigg|_{\bar\theta}
=
\begin{bmatrix}
\Pi_{k,11} & \Pi_{k,12} \\[4pt]
\Pi_{k,21} & \Pi_{k,22}
\end{bmatrix}.
\]
Since $M_{\Omega}$ decomposes block-wise as 
$\mathcal{L}_{ij} + \Sigma_{ij}$ (or $\mathcal{H}_{ij} + \Sigma_{ij}$), 
linearity of differentiation gives
\[
\Pi_{k,ij} 
= \frac{\partial \mathcal{L}_{ij}}{\partial \theta_k}\bigg|_{\bar\theta}
+ \frac{\partial \Sigma_{ij}}{\partial \theta_k}\bigg|_{\bar\theta}.
\]
Exploiting the sign convention~$\sigma$, which unifies 
$\partial_k\mathcal{L}_{ij}$ and $\partial_k\mathcal{H}_{ij}$, 
and applying the chain rule to each product of $\theta$-dependent matrices, 
the block entries expand as
\begin{align*}
\Pi_{k,11}
&= (\partial_k A)^\top P^\star + P^\star (\partial_k A)
\\
&\quad - \sigma\bigl[(\partial_k A)^\top Q^\star A + A^\top Q^\star (\partial_k A)\bigr]
\\
&\quad + \rho^\star\bigl[(\partial_k C)^\top C + C^\top (\partial_k C)\bigr],
\\[6pt]
\Pi_{k,12}
&= P^\star (\partial_k B)
- \sigma\bigl[(\partial_k A)^\top Q^\star B + A^\top Q^\star (\partial_k B)\bigr]
\\
&\quad - \tfrac{1}{2}(\partial_k C)^\top
+ \rho^\star\bigl[(\partial_k C)^\top D + C^\top (\partial_k D)\bigr],
\\[6pt]
\Pi_{k,22}
&= -\sigma\bigl[(\partial_k B)^\top Q^\star B + B^\top Q^\star (\partial_k B)\bigr]
\\
&\quad - \tfrac{1}{2}\bigl(\partial_k D + (\partial_k D)^\top\bigr)
\\
&\quad + \rho^\star\bigl[(\partial_k D)^\top D + D^\top (\partial_k D)\bigr],
\end{align*}
with $\Pi_{k,21} = \Pi_{k,12}^\top$. The passivity-index sensitivity with respect
to the $k$-th parameter follows from~\eqref{eq:envelope_gkyp} as
\begin{equation} \label{eq:sensitivities}
\frac{\partial \alpha^\star}{\partial \theta_k}\bigg|_{\bar\theta}
=
-\langle Z_M^{\star},\; \Pi_k \rangle,
\quad k = 1,\dots,p.
\end{equation}
The analytical sensitivities for the \ac{mf} case are derived analogously
and are provided in Appendix~\ref{app:mf_sensitivity}.

\section{Algorithm Design}\label{sec:bilevel}

In this section we develop a gradient-based algorithm for passivity tuning that
combines the analytical sensitivities derived from the Envelope Theorem with a
bilevel formulation that isolates the non-convexity in the outer problem.

\subsection{Bilevel Formulation}

The non-convexity of Problem~\eqref{opt:non_convex_passivity} stems exclusively from 
the coupling between the tunable parameters~$\theta$ and the matrix inequality variables
 $(P,Q,\rho,\nu)$. We hence exploit the structure and decouple 
 these by recasting the problem in bilevel form:
\begin{equation}
\label{opt:bilevel}
\begin{aligned}
\max_{\theta \in \Theta}\quad & \alpha^{\star}(\theta) \\
\text{s.t.}\quad & \alpha^{\star}(\theta) := \max_{P,\,Q,\,\rho,\,\nu}\; \alpha(\rho,\nu) \\
& \qquad\qquad\;\;\text{s.t.}\;\; Q \succeq 0,\;\; -M_{\Omega}(P,Q,\rho,\nu;\theta) \succeq 0.
\end{aligned}
\end{equation}

The inner problem solves a standard \ac{sdp} and computes the best passivity properties for a given fixed~$\theta$, which enters only as a parameter\footnote{This is indicated notation-wise by the semicolon in \eqref{opt:bilevel}.}; the outer problem searches $\Theta$ for the parameter configuration that maximizes the attained passivity index. 
This reformulation isolates the non-convexity in the outer level. In order to determine in which direction to move $\theta$ to increase $\alpha^\star$ in the outer loop, gradient information would be helpful. The sensitivities from Section~\ref{subsec:envelope_gkyp} address this by providing $\nabla_\theta \alpha^\star$, thus enabling the gradient-based bilevel scheme presented next.

\subsection{Gradient-Based Bilevel Scheme}

We collect the component-wise sensitivities from~\eqref{eq:envelope_gkyp} 
into the gradient
\[
\nabla_\theta\alpha^{\star}(\theta)
=
\begin{bmatrix}
\displaystyle\frac{\partial\alpha^{\star}}{\partial\theta_1}
& \cdots &
\displaystyle\frac{\partial\alpha^{\star}}{\partial\theta_p}
\end{bmatrix}^{\!\top}\!,
\]
which drives a projected gradient ascent on the outer problem~\eqref{opt:bilevel}. Each iteration~$k$ performs three steps: 
solve the inner \ac{sdp} at~$\theta^{(k)}$, evaluate $\nabla_\theta\alpha^{\star}(\theta^{(k)})$ in~\eqref{eq:sensitivities}, and update
\begin{equation*}
\theta^{(k+1)} = \theta^{(k)} + d_k,
\end{equation*}
where
\begin{equation*}
d_k := \mathcal{P}_{\Theta}\!\bigl(
\theta^{(k)} + \eta_k\,\nabla_\theta\alpha^{\star}(\theta^{(k)})
\bigr) - \theta^{(k)}
\end{equation*}
denotes the projected displacement, $\eta_k > 0$ is the step size at 
iteration $k$, and $\mathcal{P}_{\Theta} \colon \R^p \to \Theta$ is the 
Euclidean projection onto the feasible set~$\Theta$.

Three remaining design choices complete the scheme: the step-size 
selection rule for $\eta_k$, the stopping criterion and a stability safeguard.

\paragraph{Step-size selection}

We select $\eta_k$ via Armijo backtracking adapted to projected 
steps~\cite{nocedal2006numerical}. Starting from an initial 
trial $\eta > 0$, the algorithm contracts by a factor $\beta \in (0,1)$ 
until
\begin{equation}
\label{eq:projected_armijo}
\alpha^{\star}\!\bigl(\theta^{(k)} + d_k\bigr)
\;\geq\;
\alpha^{\star}(\theta^{(k)})
\;+\;
c\,\nabla_\theta\alpha^{\star}(\theta^{(k)})^{\!\top} d_k
\end{equation}
holds, with $c \in (0,1)$.

Every accepted nonzero $d_k$ therefore produces a strict increase in $\alpha^\star$ and preserves feasibility. The parameters \(\beta\) and \(c\) must be chosen by the user and affect the convergence behavior of the algorithm.

\paragraph{Stopping criterion}
Because the maximization is constrained to~$\Theta$, we
monitor the norm of the projected gradient 
mapping~\cite{bertsekas1999nonlinear}
\begin{equation}
\label{eq:grad_mapping}
G^{(k)}
:=
\frac{1}{\bar{\eta}}
\Bigl\|
\mathcal{P}_{\Theta}\!\bigl(
\theta^{(k)} + \bar{\eta}\,\nabla_\theta\alpha^{\star}(\theta^{(k)})
\bigr) - \theta^{(k)}
\Bigr\|,
\end{equation}
evaluated with a \emph{fixed} reference step $\bar{\eta} > 0$, which we set 
equal to the initial Armijo trial step~$\eta_0$. The quantity~$G^{(k)}$ 
reduces to $\|\nabla_\theta\alpha^{\star}(\theta^{(k)})\|$ when no constraint 
is active. The iteration is stopped when
\begin{equation}
\label{eq:stopping}
G^{(k)} < \varepsilon_g ,
\end{equation}
where $\varepsilon_g > 0$ is a user-defined numerical tolerance.
\paragraph{Stability safeguard}
Passivity characterizes input--output behavior; it does not, by itself, guarantee Lyapunov stability.
In particular, a system with negative passivity indices in some frequency range may satisfy the \ac{gkyp} \ac{lmi}~\eqref{eq:M_omega}
while being unstable. To prevent this, we augment the inner \ac{sdp} with an additional constraint
\begin{equation}
\label{eq:stab_lmi}
A(\theta)^{\!\top} P_s + P_s\,A(\theta)
\prec 0,
\qquad
P_s \succ 0,
\end{equation}
where $P_s \in \mathbb{S}^n$ is an additional decision variable.
Constraint~\eqref{eq:stab_lmi} certifies that the optimized system remains Lyapunov stable at every iterate.

Let~$Z_s^{\star} \succeq 0$ denote the optimal dual multiplier of~\eqref{eq:stab_lmi}. 
This constraint contributes
\begin{equation}
\label{eq:stab_grad}
-\bigl\langle Z_s^{\star},\;
(\partial_k A)^{\!\top} P_s^{\star} + P_s^{\star}\,(\partial_k A)
\bigr\rangle,
\qquad k = 1,\ldots,p,
\end{equation}
to the $k$-th component of~$\nabla_\theta\alpha^{\star}$. 
The line search accepts an update only if~\eqref{eq:stab_lmi} remains feasible, 
thus preserving closed-loop stability at every accepted iterate.

Algorithm~\ref{alg:bilevel} summarizes the complete procedure.

\begin{remark}[Nonsmooth case]
\label{rem:nonsmooth}
If Assumption~\ref{ass:local_regularity}~(iii) fails (see Remark~\ref{rem:directional_derivative}), the right-hand side of~\eqref{eq:envelope_gradient} remains computable and serves as a candidate search direction. The Armijo condition~\eqref{eq:projected_armijo} of our algorithm guarantees ascent for every accepted nonzero step, also in the nonsmooth case.
\end{remark}

\begin{algorithm}[t]
\caption{Projected Gradient Ascent for Passivity-Index Optimization}
\label{alg:bilevel}
\begin{algorithmic}
    \REQUIRE Initial point $\theta^{(0)} \in \Theta$;\; reference step $\bar{\eta} > 0$;\; tolerance $\varepsilon_g > 0$
    \STATE $k \gets 0$
    \REPEAT
        \STATE Solve the inner \ac{sdp} at~$\theta^{(k)}$, augmented with~\eqref{eq:stab_lmi}
        \STATE Retrieve dual multipliers $Z_M^{\star}$ and $Z_s^{\star}$
        \STATE Evaluate $\nabla_\theta\alpha^{\star}(\theta^{(k)})$ from~\eqref{eq:envelope_gkyp} and~\eqref{eq:stab_grad}
        \STATE Evaluate the gradient mapping $G^{(k)}$ from~\eqref{eq:grad_mapping}
        \STATE Backtrack $\eta_k$ until $d_k \gets \mathcal{P}_{\Theta}\!\bigl(\theta^{(k)} + \eta_k\,\nabla_\theta\alpha^{\star}(\theta^{(k)})\bigr) - \theta^{(k)}$ satisfies~\eqref{eq:projected_armijo}, else $d_k \gets 0$
        \STATE $\theta^{(k+1)} \gets \theta^{(k)} + d_k$
        \STATE $k \gets k+1$
    \UNTIL{$G^{(k-1)} < \varepsilon_g$}
    \RETURN $\theta^{(k)}$
\end{algorithmic}
\end{algorithm}
\section{Numerical Simulations}\label{sec:simulations}
 
We evaluate the proposed tuning method on two systems of increasing complexity and scale: a grid-following converter and a multi-converter microgrid. The first case considers the problem of enlarging the passive region of a grid-following converter, which is a well-known problem in the literature and is currently solved by trial and error~\cite{Chen2024}.
The second case shows the same framework improving the aggregate passivity margin seen from the \ac{pcc} of an interconnected converter network in a specific \ac{mf} range.

\subsection{Common Algorithmic Setup}\label{sec:common_setup}
The tunable parameters, i.e., the optimization variables, are initialized at their nominal values. Furthermore, 
we impose constraints of $\pm 50\%$ on each parameter. These bounds serve demonstration purposes only; practitioners
may define them individually for each parameter in practical applications. All runs employ the Armijo backtracking rule (Section~\ref{sec:bilevel}) and converge when the projected gradient mapping~\eqref{eq:grad_mapping}, evaluated with $\bar{\eta} = \eta_0$, drops below the tolerance~$\varepsilon_g$. Tables~\ref{tab:hyperparams_cc} and~\ref{tab:hyperparams_mc} report the hyperparameters for each case study. In both case studies, the tunable gains and parameters span several orders of magnitude ($10^{-2}$ to $10^{3}$), which ill-conditions the gradient-ascent problem. We mitigate this by optimizing in log-space~\cite{kivinen1997exponentiated}. We implement the symbolic linearization and all numerical experiments in Julia~\cite{bezanson2017julia}.

\subsection{Case~1: Grid-Following Converter}\label{sec:case_cc}
 
\subsubsection{System Description}
 
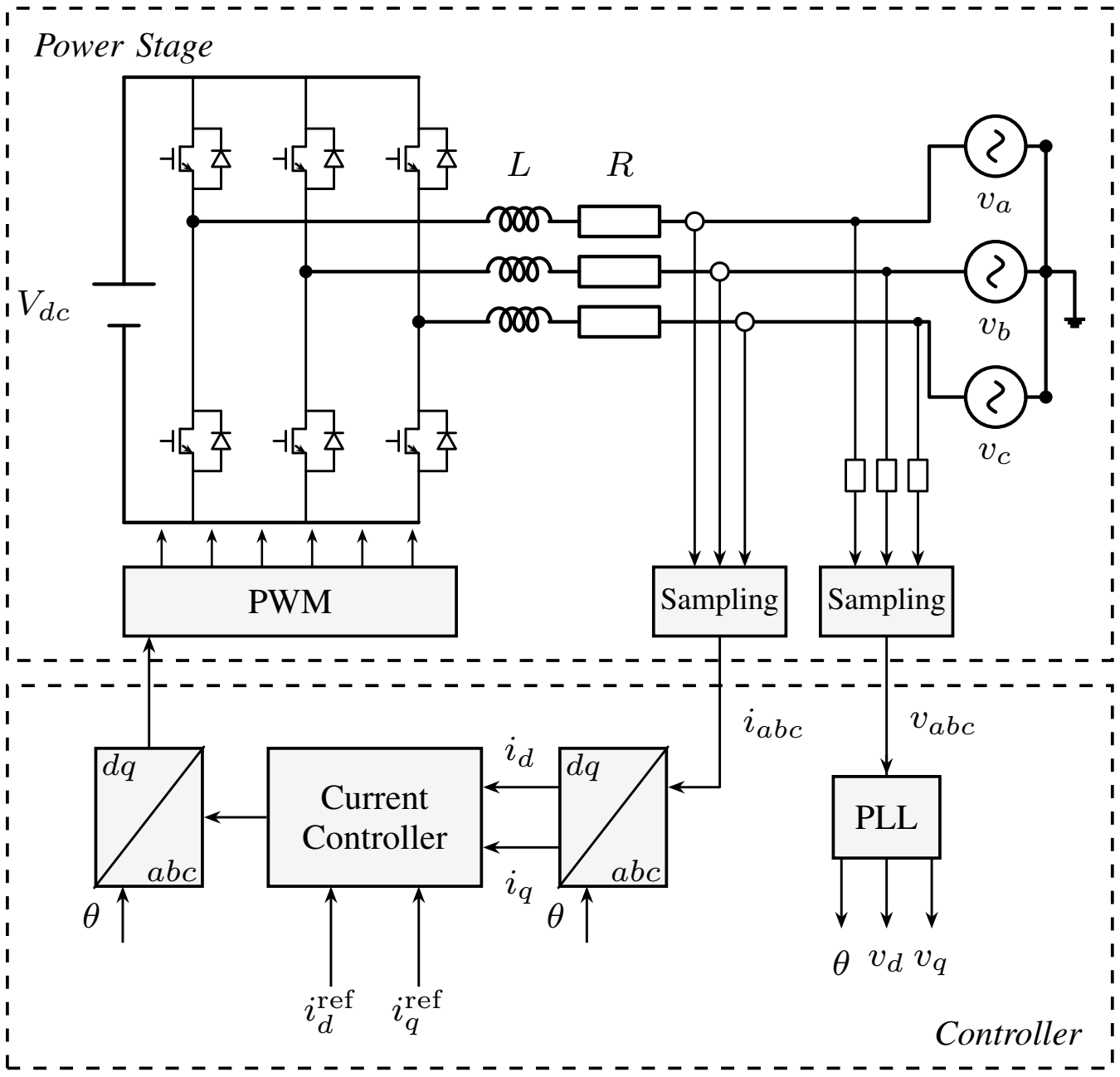
\begin{figure}[!t]
\centering
\begingroup
\ctikzset{bipoles/thickness=1, inductors/coils=4, inductor=cute,
          bipoles/length=0.8cm}%
\begin{tikzpicture}[
  x=1cm,y=1cm,
  font=\footnotesize,
  pwire/.style={line width=0.8pt, black, line cap=round, line join=round},
  swire/.style={line width=0.55pt, black, line cap=round, line join=round},
  thinw/.style={line width=0.55pt, black, line cap=round, line join=round},
  sarr/.style={-{Stealth[length=3.8pt]}, line width=0.55pt, black},
  garr/.style={-{Stealth[length=3.2pt]}, line width=0.5pt, black},
  blk/.style={draw, line width=0.7pt, inner sep=2.5pt, align=center, fill=gray!8},
  tap/.style={circle, fill=black, inner sep=0pt, minimum size=3pt},
  sensor/.style={circle, draw, line width=0.6pt, fill=white, inner sep=0pt, minimum size=4pt},
]

\newcommand{\igbtsw}[1]{%
  \begin{scope}[shift={(#1)}, scale=0.8]
    \draw[thinw] (0,0.30) -- (0,0.12);
    \draw[thinw] (0,-0.12) -- (0,-0.30);
    \draw[thinw] (-0.13,0.12) -- (0,0.12);
    \draw[thinw] (-0.13,-0.12) -- (0,-0.12);
    \draw[thinw] (-0.13,0.12) -- (-0.13,-0.12);
    \draw[thinw] (-0.20,-0.10) -- (-0.20,0.10);
    \draw[thinw] (-0.32,0) -- (-0.20,0);
    \draw[thinw,-{Stealth[length=2.2pt]}] (-0.10,-0.05) -- (0,-0.12);
    \draw[thinw] (0.28,0.30) -- (0.28,0.08);
    \draw[thinw] (0.28,-0.30) -- (0.28,-0.08);
    \draw[thinw] (0.19,-0.08) -- (0.37,-0.08) -- (0.28,0.08) -- cycle;
    \draw[thinw] (0.19,0.08) -- (0.37,0.08);
    \draw[thinw] (0,0.30) -- (0.28,0.30);
    \draw[thinw] (0,-0.30) -- (0.28,-0.30);
  \end{scope}
}

\def\railT{4.60}\def\railB{1.05}
\def\ya{3.45}\def\yb{3.05}\def\yc{2.65}      
\def\sa{4.05}\def\sb{3.05}\def\sc{2.05}      

\draw[pwire] (1.00,\railT) -- (1.00,2.95);
\draw[pwire] (1.00,2.62) -- (1.00,\railB);
\draw[pwire] (0.76,2.95) -- (1.24,2.95);
\draw[pwire] (0.88,2.62) -- (1.12,2.62);
\node[anchor=east] at (0.71,2.78) {$V_{dc}$};

\draw[pwire] (1.00,\railT) -- (3.35,\railT);
\draw[pwire] (1.00,\railB) -- (3.35,\railB);

\foreach \x in {1.55,2.45,3.35}{
  \draw[thinw] (\x,\railT) -- (\x,4.19);
  \igbtsw{(\x,3.95)}
  \draw[thinw] (\x,3.71) -- (\x,1.94);
  \igbtsw{(\x,1.70)}
  \draw[thinw] (\x,1.46) -- (\x,\railB);
}
\node[tap] at (1.55,\ya) {};
\node[tap] at (2.45,\yb) {};
\node[tap] at (3.35,\yc) {};

\draw[pwire] (1.55,\ya) -- (3.75,\ya) to[L] (4.55,\ya) to[R] (5.35,\ya)
  -- (7.42,\ya) -- (7.42,\sa) -- (7.55,\sa) to[sV] (8.35,\sa);
\draw[pwire] (2.45,\yb) -- (3.75,\yb) to[L] (4.55,\yb) to[R] (5.35,\yb)
  -- (7.55,\sb) to[sV] (8.35,\sb);
\draw[pwire] (3.35,\yc) -- (3.75,\yc) to[L] (4.55,\yc) to[R] (5.35,\yc)
  -- (7.42,\yc) -- (7.42,\sc) -- (7.55,\sc) to[sV] (8.35,\sc);
\node[anchor=south] at (4.15,\ya+0.24) {$L$};
\node[anchor=south] at (4.95,\ya+0.24) {$R$};
\node[anchor=north] at (7.95,\sa-0.25) {$v_a$};
\node[anchor=north] at (7.95,\sb-0.25) {$v_b$};
\node[anchor=north] at (7.95,\sc-0.25) {$v_c$};
\draw[pwire] (8.35,\sa) -- (8.35,\sc);
\node[tap] at (8.35,\sa) {};
\node[tap] at (8.35,\sb) {};
\node[tap] at (8.35,\sc) {};
\draw[pwire] (8.35,\sb) -- (8.58,\sb) -- (8.58,\sb-0.38);
\node[tlground, scale=0.7] at (8.58,\sb-0.38) {};

\node[blk, minimum width=1.0cm, minimum height=0.55cm, font=\scriptsize, inner sep=1.6pt] (samp1) at (5.75,0.42) {Sampling};
\foreach \x/\y in {5.55/\ya, 5.75/\yb, 5.95/\yc}{
  \node[sensor] at (\x,\y) {};
  \draw[sarr] (\x,\y-0.07) -- (\x,0.695);
}

\node[blk, minimum width=1.0cm, minimum height=0.55cm, font=\scriptsize, inner sep=1.6pt] (samp2) at (7.08,0.42) {Sampling};
\foreach \x/\y in {6.83/\ya, 7.08/\yb, 7.33/\yc}{
  \draw[swire] (\x,\y) -- (\x,1.55);
  \draw[swire, fill=white] (\x-0.075,1.30) rectangle (\x+0.075,1.55);
  \draw[sarr] (\x,1.30) -- (\x,0.695);
  \node[tap, minimum size=2.2pt] at (\x,\y) {};
}

\node[blk, minimum width=2.65cm, minimum height=0.55cm] (pwm) at (2.325,0.42) {PWM};
\foreach \x in {1.30,1.70,2.10,2.50,2.90,3.30}{
  \draw[garr] (\x,0.695) -- (\x,1.00);
}

\draw[dashed, line width=0.7pt] (0.07,-0.05) rectangle (8.88,5.15);
\node[anchor=north west, font=\footnotesize\itshape] at (0.16,5.08) {Power Stage};

\node[blk, minimum width=0.85cm, minimum height=1.1cm] (mtr) at (1.20,-1.30) {};
\draw[thinw] (mtr.south west) -- (mtr.north east);
\node[anchor=north west, inner sep=1.4pt, font=\scriptsize] at (mtr.north west) {$dq$};
\node[anchor=south east, inner sep=1.4pt, font=\scriptsize] at (mtr.south east) {$abc$};
\draw[sarr] (mtr.north) -- (1.20,0.145);
\draw[sarr] (0.99,-2.30) -- (0.99,-1.85);
\node[anchor=east] at (0.94,-2.08) {$\theta$};

\node[blk, minimum width=1.7cm, minimum height=1.1cm] (cc) at (3.00,-1.30) {Current\\Controller};
\draw[sarr] (cc.west) -- (mtr.east);

\node[blk, minimum width=0.85cm, minimum height=1.1cm] (mea) at (4.90,-1.30) {};
\draw[thinw] (mea.south west) -- (mea.north east);
\node[anchor=north west, inner sep=1.4pt, font=\scriptsize] at (mea.north west) {$dq$};
\node[anchor=south east, inner sep=1.4pt, font=\scriptsize] at (mea.south east) {$abc$};
\draw[sarr] (4.69,-2.30) -- (4.69,-1.85);
\node[anchor=east] at (4.64,-2.08) {$\theta$};

\draw[sarr] ($(mea.west)+(0,0.24)$) -- ($(cc.east)+(0,0.24)$);
\draw[sarr] ($(mea.west)+(0,-0.24)$) -- ($(cc.east)+(0,-0.24)$);
\node[anchor=south] at (4.16,-1.02) {$i_d$};
\node[anchor=north] at (4.16,-1.60) {$i_q$};

\draw[swire] (samp1.south) -- (5.75,-1.06);
\draw[sarr]  (5.75,-1.06) -- ($(mea.east)+(0,0.24)$);
\node[anchor=west] at (5.81,-0.55) {$i_{abc}$};

\draw[sarr] (2.65,-2.65) -- (2.65,-1.85);
\draw[sarr] (3.35,-2.65) -- (3.35,-1.85);
\node[anchor=north] at (2.65,-2.55) {$i_d^{\mathrm{ref}}$};
\node[anchor=north] at (3.35,-2.55) {$i_q^{\mathrm{ref}}$};

\node[blk, minimum width=0.85cm, minimum height=0.65cm] (pll) at (7.08,-1.30) {PLL};
\draw[swire] (samp2.south) -- (7.08,-0.975);
\draw[sarr]  (7.08,-0.80) -- (pll.north);
\node[anchor=west] at (7.14,-0.55) {$v_{abc}$};
\foreach \dx/\lab in {-0.36/{\theta}, 0/{v_d}, 0.36/{v_q}}{
  \draw[sarr] ($(pll.south)+(\dx,0)$) -- ++(0,-0.55);
  \node[anchor=north] at ($(pll.south)+(\dx,-0.60)$) {$\lab$};
}

\draw[dashed, line width=0.7pt] (0.07,-3.30) rectangle (8.88,-0.25);
\node[anchor=south east, font=\footnotesize\itshape] at (8.76,-3.24) {Controller};

\end{tikzpicture}%
\endgroup
\caption{Grid-following \ac{cc} converter with \ac{pwm}
delay in the $dq$ reference frame.}
\label{fig:cc_converter}
\end{figure}

The system is a three-phase \ac{cc} converter operating in the $dq$ synchronous reference frame, shown in Fig.~\ref{fig:cc_converter}. Its control architecture comprises a \ac{pi} current controller (one per axis), a \ac{pll}, and a first-order approximation of the \ac{pwm} delay.  The model and nominal parameters follow the case study of~\cite[Section~VI, Table~III]{Wen2016}. Two simplifications are introduced with respect to the original
benchmark:
\begin{enumerate}
\item[(i)] The signal-conditioning filter is omitted.
\item[(ii)] The time delay is approximated by the
      first-order transfer function $1/(1+sT_d)$.
\end{enumerate}
 
Table~\ref{tab:plant_params} lists the system parameters. The switching frequency is $f_s = 5$\,kHz, yielding a computational
delay $T_d = 1.5/f_s = 300$\,$\mu$s.  The system is linearized around an equilibrium corresponding to current injection mode,
producing a state-space realization $(A,B,C,D) \in \R^{8\times 8}\times\R^{8\times 2} \times\R^{2\times 8}\times\R^{2\times 2}$, with inputs $(V_d, V_q)$, the $dq$ components of the grid voltage at the converter terminals, and outputs $(i_d, i_q)$, the $dq$ components of the current injected into the grid through the filter inductor~$L$. This input--output pair defines the admittance port model of the converter, at which we evaluate passivity.

\begin{table}[!t]
\centering
\caption{System parameters for the grid-following converter.}
\label{tab:plant_params}
\begin{tabular}{lll}
\toprule
Symbol & Description & Value \\
\midrule
$V_{\mathrm{dc}}$       & DC-link voltage            & $600$\,V          \\
$V_d$                    & $d$-axis grid voltage      & $293$\,V          \\
$V_q$                    & $q$-axis grid voltage      & $0$\,V            \\
$L$                      & Filter inductance          & $1$\,mH           \\
$R$                      & Filter resistance          & $10$\,m$\Omega$   \\
$f_s$                    & Switching frequency        & $5$\,kHz          \\
$T_d$                    & PWM delay                  & $300$\,$\mu$s     \\
$\omega_g$               & Grid frequency             & $2\pi\cdot50$\,rad/s \\
$i_d^{\mathrm{ref}}$     & $d$-axis current reference & $190$\,A          \\
$i_q^{\mathrm{ref}}$     & $q$-axis current reference & $0$\,A            \\
\bottomrule
\end{tabular}
\end{table}

\subsubsection{Tunable Parameters and Bounds}
 
The tunable parameter vector collects the \ac{pi} gains of the \acp{cc} in $d$ and $q$ and the \ac{pll}:
\[
\theta = \bigl[\,k_p^{d}\,;\; k_i^{d}\,;\; k_p^{q}\,;\; k_i^{q}\,;\;
k_p^{\mathrm{PLL}}\,;\; k_i^{\mathrm{PLL}}\bigr]^\top\!,
\]
initialized at the nominal values from~\cite{Wen2016}:
\[
\theta^{(0)} = \bigl[\,0.0105\,;\; 1.1519\,;\; 0.0105\,;\; 1.1519\,;\;
3.0\,;\; 3.2\,\bigr]^\top\!.
\]
 
The objective is to maximize the \ac{ifp} index $\nu$ (i.e. $\bar{\rho}=0$) over
\[
    \Omega=[\omega_1, +\infty)=[500,+\infty)\ \mathrm{rad/s}.
\] 
With this setup, we tune the parameters to maximize passivity in the high-frequency range above~$\omega_1$ 
(see the gray shaded region in Fig.~\ref{fig:ifp_frequency}), 
which is a common problem faced in grid-tied
\acp{vsc}~\cite{Chen2024}.
The resulting
bounds and hyperparameters are given in Tables~\ref{tab:bounds}
and~\ref{tab:hyperparams_cc}.
 
\begin{table}[!t]
\centering
\caption{Box constraints on the tunable parameters (Case~1).}
\label{tab:bounds}
\begin{tabular}{lc}
\toprule
Parameter & $\Theta_j$ \\
\midrule
$k_p^{d}$              & $[0.0053,\; 0.0158]$ \\
$k_i^{d}$              & $[0.576,\; 1.728]$   \\
$k_p^{q}$              & $[0.0053,\; 0.0158]$ \\
$k_i^{q}$              & $[0.576,\; 1.728]$   \\
$k_p^{\mathrm{PLL}}$   & $[1.5,\; 4.5]$   \\
$k_i^{\mathrm{PLL}}$   & $[1.6,\; 4.8]$   \\
\bottomrule
\end{tabular}
\end{table}
 
\begin{table}[!t]
\centering
\caption{Algorithm hyperparameters (Case~1).}
\label{tab:hyperparams_cc}
\begin{tabular}{lll}
\toprule
Symbol & Description & Value \\
\midrule
$\eta_0$       & Initial step size    & $0.05$    \\
$\beta$        & Backtracking factor  & $0.5$     \\
$c$            & Armijo constant      & $0.1$     \\
               & Maximum backtracks   & $2$       \\
$\varepsilon_g$ & Stationarity tolerance & $10^{-2}$ \\
\bottomrule
\end{tabular}
\end{table}

\subsubsection{Gradient Validation}
 
Before launching the optimization, we compare the analytical
gradient derived via the Envelope Theorem against central \ac{fd}~\cite{nocedal2006numerical} at~$\theta^{(0)}$.  For
each component~$j$, the central-difference approximation takes
the form
\[
\frac{\partial \nu}{\partial \theta_j}
\approx
\frac{\nu(\theta + \delta\, e_j) - \nu(\theta - \delta\, e_j)}{2\delta},
\]
where~$e_j$ denotes the~$j$th canonical basis vector and
$\delta = 10^{-5}$ is the perturbation step size.
 
We evaluate both gradients at~$\theta^{(0)}$, which yields
$\nu(\theta^{(0)}) = -0.395$ in the frequency range of interest (see Fig.~\ref{fig:ifp_frequency}).
Table~\ref{tab:fd_gradient_check} reports the results
component-wise. The two
gradients match closely, confirming the analytical gradient derivation in~\eqref{eq:envelope_gkyp}.
 
\begin{table}[!t]
\centering
\caption{Central finite-difference validation of the envelope theorem
gradient at $\theta^{(0)}$ ($\delta=10^{-5}$).}
\label{tab:fd_gradient_check}
\scriptsize
\setlength{\tabcolsep}{3pt}
\begin{tabular}{lrr}
\toprule
Component & Envelope & Central FD \\
\midrule
$\partial\nu/\partial k_p^{d}$             & $6.21\times10^{-2}$  & $6.21\times10^{-2}$  \\
$\partial\nu/\partial k_i^{d}$             & $2.41\times10^{-4}$  & $2.41\times10^{-4}$  \\
$\partial\nu/\partial k_p^{q}$             & $-7.49\times10^{0}$  & $-7.49\times10^{0}$  \\
$\partial\nu/\partial k_i^{q}$             & $-5.46\times10^{-2}$ & $-5.46\times10^{-2}$ \\
$\partial\nu/\partial k_p^{\mathrm{PLL}}$  & $-1.28\times10^{-1}$ & $-1.28\times10^{-1}$ \\
$\partial\nu/\partial k_i^{\mathrm{PLL}}$  & $-1.25\times10^{-4}$ & $-1.25\times10^{-4}$ \\
\bottomrule
\end{tabular}
\end{table}

The computational cost advantage is direct: no extra \ac{sdp} solves beyond the one
required for the objective evaluation, while \ac{fd}
gradients require additional solves that grow linearly in the
number of tuning parameters~$p$: central \ac{fd} requires $2p$ solves,
forward/backward \ac{fd} requires~$p$
(see Table~\ref{tab:grad_solve_cost}).
 
\begin{table}[!t]
\centering
\caption{Additional SDP solves per gradient evaluation.}
\label{tab:grad_solve_cost}
\scriptsize
\begin{tabular}{lcc}
\toprule
Method & Extra solves & For $p=6$ \\
\midrule
Envelope gradient (proposed) & $0$  & $0$  \\
Forward/backward FD          & $p$  & $6$  \\
Central FD                   & $2p$ & $12$ \\
\bottomrule
\end{tabular}
\end{table}

\subsubsection{Results}\label{sec:cc_results}
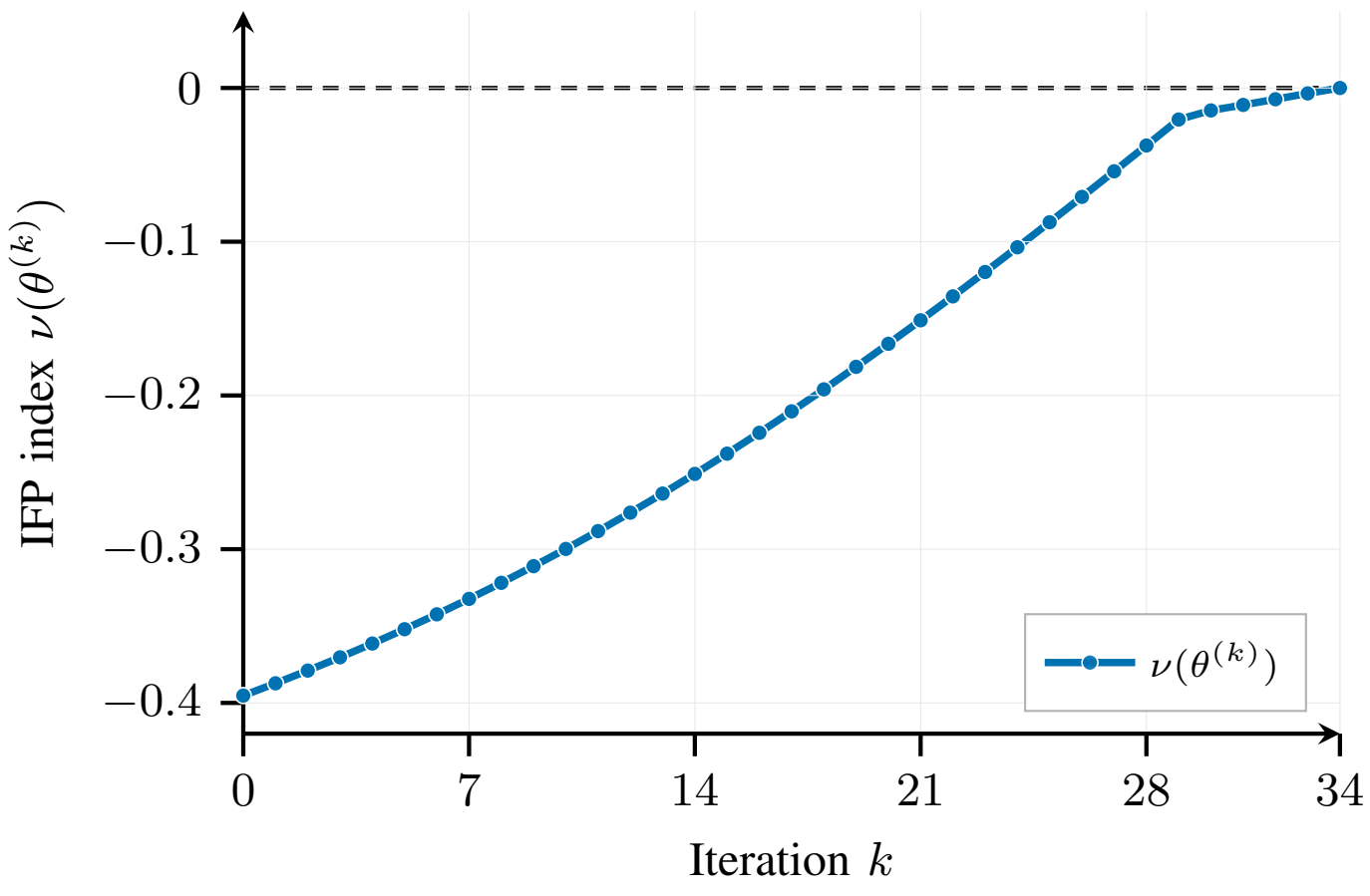
\begin{figure}[!t]
\centering
\begin{tikzpicture}
\definecolor{optcol}{RGB}{0,114,178}    
\definecolor{threshcol}{RGB}{118,42,131} 

\begin{axis}[
    width=\columnwidth,
    height=0.72\columnwidth,
    xlabel={Iteration $k$},
    ylabel={IFP index $\nu(\theta^{(k)})$},
    xmin=0, xmax=34,
    ymin=-0.42, ymax=0.05,
    xtick={0,7,14,21,28,34},
    ytick={-0.4,-0.3,-0.2,-0.1,0},
    grid=major,
    grid style={line width=0.1pt, draw=gray!12},
    major grid style={line width=0.2pt, draw=gray!18},
    axis lines=left,
    axis line style={-stealth, thick},
    tick align=outside,
    tick style={black, thick},
    label style={font=\small},
    tick label style={font=\small},
    enlargelimits=false,
    clip=true,
    axis on top=true,
    legend pos=south east,
    legend cell align=left,
    legend style={
        draw=gray!60,
        fill=white,
        fill opacity=0.92,
        text opacity=1,
        font=\scriptsize,
        inner sep=3pt,
        row sep=1pt,
        cells={anchor=west}
    }
]

\addplot[black, densely dashed, line width=0.8pt, forget plot]
coordinates {(0,0) (34,0)};

\addplot[optcol, line width=1.4pt, mark=*, mark size=1.5pt,
         mark options={fill=optcol, draw=white, line width=0.3pt}]
table[x=iteration, y=nu, col sep=comma]
{figures_data/cc_obj.csv};
\addlegendentry{$\nu(\theta^{(k)})$}

\end{axis}
\end{tikzpicture}
\caption{Minimum \ac{ifp} index~$\nu$
over the frequency range of interest of the \ac{cc} converter across the optimization iterations.}
\label{fig:obj_evolution}
\end{figure}

Figure~\ref{fig:obj_evolution} shows the minimum \ac{ifp} index over the
frequency range of interest across the optimization iterations. The algorithm
converges in 34 iterations, increasing the index
from $\nu(\theta^{(0)}) = -0.395$ to $\nu(\theta^{(34)})\approx -1.01\times10^{-5}$.
Hence, the passivity
shortage over $[500,+\infty)$\,rad/s is reduced to numerical
tolerance.
 
The tuned gains are reported in Table~\ref{tab:tuning_results}.
The dominant changes occur in the \ac{pll} proportional gain and in
the $q$-axis current controller. Specifically,
$k_p^{\mathrm{PLL}}$ decreases by $50\%$, $k_p^q$ decreases by
$24\%$, and $k_i^q$ decreases by $12\%$. The remaining gains change
by at most $0.3\%$.
\begin{table}[!t]
\centering
\caption{Tuning results: initial versus tuned gains.
Parameters marked with $\dagger$ saturate a bound.}
\label{tab:tuning_results}
\begin{tabular}{lccc}
\toprule
Parameter & $\theta_j^{(0)}$ & $\theta_j^{(34)}$ & Change \\
\midrule
$k_p^{d}$              & $0.0105$  & $0.01053$ & $+0.3\,\%$   \\
$k_i^{d}$              & $1.1519$  & $1.1523$  & $+0.03\,\%$  \\
$k_p^{q}$              & $0.0105$  & $0.00796$ & $-24.2\,\%$  \\
$k_i^{q}$              & $1.1519$  & $1.0135$  & $-12.0\,\%$  \\
$k_p^{\mathrm{PLL}}{}^{\dagger}$   & $3.0$     & $1.5$     & $-50\,\%$    \\
$k_i^{\mathrm{PLL}}$   & $3.2$     & $3.1997$  & $-0.009\,\%$ \\
\bottomrule
\end{tabular}
\end{table}

\begin{figure}[!t]
\centering
\begin{tikzpicture}
\definecolor{cA}{RGB}{213,94,0}    
\definecolor{cB}{RGB}{0,114,178}   
\definecolor{cC}{RGB}{0,158,115}   
\definecolor{cD}{RGB}{204,121,167} 
\definecolor{cE}{RGB}{230,159,0}   
\definecolor{cF}{RGB}{86,180,233}  

\begin{axis}[
    width=\columnwidth,
    height=0.72\columnwidth,
    xlabel={Iteration $k$},
    ylabel={$\Delta_j^{(k)}$ [\%]},
    xmin=0, xmax=34,
    ymin=-55, ymax=8,
    xtick={0,7,14,21,28,34},
    ytick={-50,-40,-30,-20,-10,0},
    grid=major,
    grid style={line width=0.1pt, draw=gray!12},
    major grid style={line width=0.2pt, draw=gray!18},
    axis lines=left,
    axis line style={-stealth, thick},
    tick align=outside,
    tick style={black, thick},
    label style={font=\small},
    tick label style={font=\small},
    enlargelimits=false,
    clip=true,
    axis on top=true,
    legend columns=3,
    legend style={
        at={(0.5,1.15)},
        anchor=south,
        draw=gray!60,
        fill=white,
        font=\scriptsize,
        nodes={scale=1.0, transform shape},
        legend image post style={scale=0.7},
        inner sep=2pt,
        column sep=4pt,
        row sep=1pt,
        /tikz/every even column/.append style={column sep=6pt}
    }
]

\addplot[black, densely dashed, line width=0.8pt, forget plot]
coordinates {(0,0) (34,0)};

\addplot[cB, line width=1.1pt, mark=*, mark size=1.5pt, mark repeat=1,
         mark options={fill=cB, draw=white, line width=0.2pt}]
table[x=iteration, y=kp_CC_d_pct, col sep=comma]
{figures_data/cc_par.csv};
\addlegendentry{$k_p^d$}

\addplot[cE, densely dashed, line width=1.1pt, mark=diamond*, mark size=1.6pt, mark repeat=1,
         mark options={solid, fill=cE, draw=white, line width=0.2pt}]
table[x=iteration, y=ki_CC_d_pct, col sep=comma]
{figures_data/cc_par.csv};
\addlegendentry{$k_i^d$}

\addplot[cC, line width=1.3pt, mark=square*, mark size=1.5pt, mark repeat=1,
         mark options={fill=cC, draw=white, line width=0.2pt}]
table[x=iteration, y=kp_CC_q_pct, col sep=comma]
{figures_data/cc_par.csv};
\addlegendentry{$k_p^q$}

\addplot[cD, line width=1.3pt, mark=triangle*, mark size=1.7pt, mark repeat=1,
         mark options={fill=cD, draw=white, line width=0.2pt}]
table[x=iteration, y=ki_CC_q_pct, col sep=comma]
{figures_data/cc_par.csv};
\addlegendentry{$k_i^q$}

\addplot[cA, line width=1.4pt, mark=*, mark size=1.6pt, mark repeat=1,
         mark options={fill=cA, draw=white, line width=0.2pt}]
table[x=iteration, y=kp_PLL_pct, col sep=comma]
{figures_data/cc_par.csv};
\addlegendentry{$k_p^{PLL}$}

\addplot[cF, densely dashed, line width=1.1pt, mark=x, mark size=1.8pt, mark repeat=1,
         mark options={solid, draw=cF, line width=0.7pt}]
table[x=iteration, y=ki_PLL_pct, col sep=comma]
{figures_data/cc_par.csv};
\addlegendentry{$k_i^{PLL}$}

\end{axis}
\end{tikzpicture}
\caption{Relative change~$\Delta_j^{(k)}$ [\%] in \ac{cc} converter parameters
at each optimization iteration.}
\label{fig:parameter_history}
\end{figure}
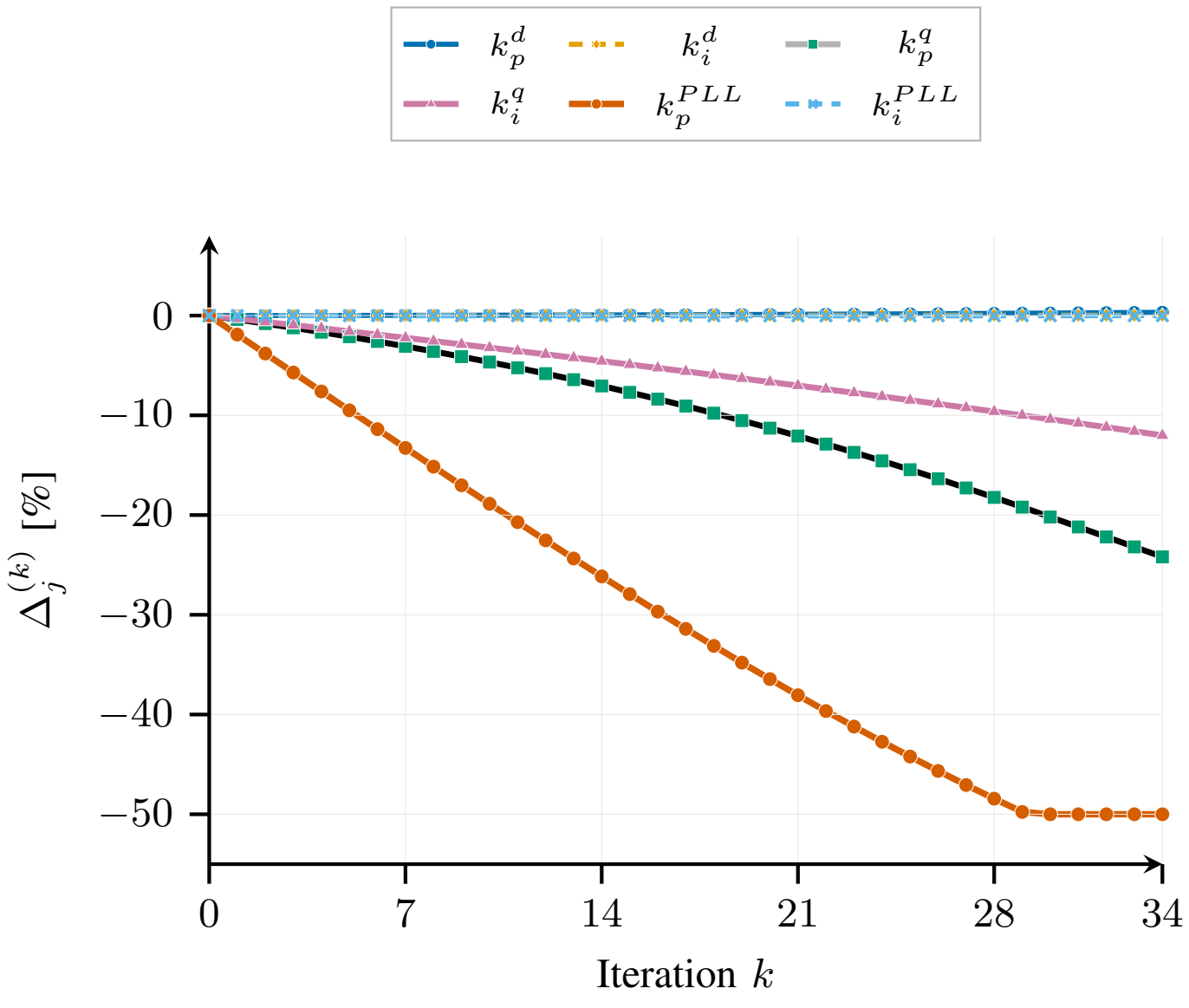

Figure~\ref{fig:parameter_history} displays the relative change
\(
  \Delta_j^{(k)}=
  \frac{\theta_j^{(k)} - \theta_j^{(0)}}{\theta_j^{(0)}}\,[\%]
\)
of each gain across iterations. Only three of the six parameters move, while the others remain essentially at their nominal values: the proposed algorithm thus automatically identifies which gains shape passivity properties in the target \ac{hf} range.

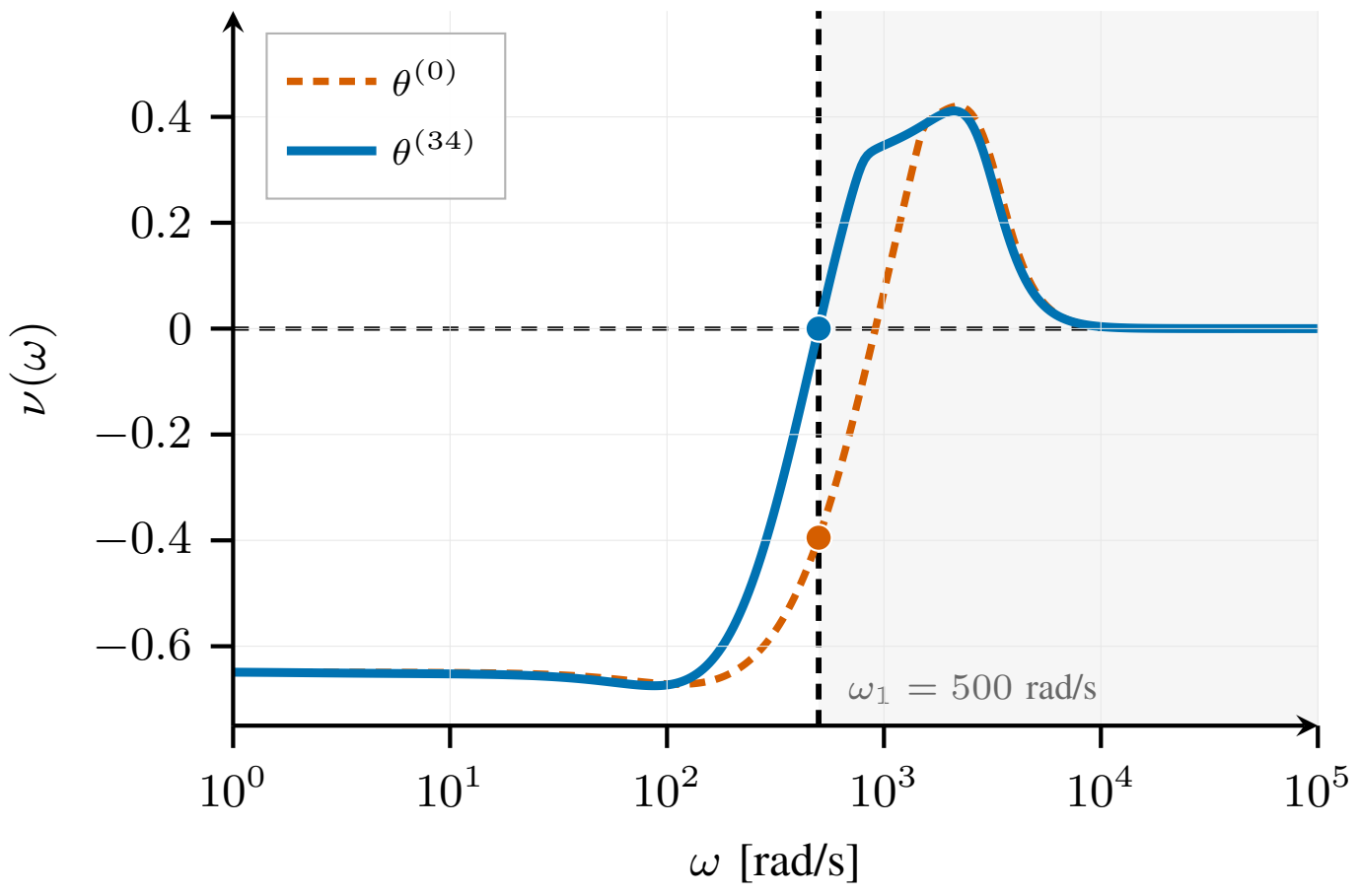
\begin{figure}[!t]
\centering
\begin{tikzpicture}
\definecolor{initcol}{RGB}{213,94,0}    
\definecolor{optcol}{RGB}{0,114,178}    
\definecolor{bandcol}{RGB}{0,114,178}   
\definecolor{infcol}{RGB}{230,159,0}    
\definecolor{threshcol}{RGB}{118,42,131}  

\begin{axis}[
    width=\columnwidth,
    height=0.72\columnwidth,
    xmode=log,
    xlabel={$\omega$ [rad/s]},
    ylabel={$\nu(\omega)$},
    xmin=1e0, xmax=1e5,
    ymin=-0.75, ymax=0.6,
    xtick={1e0, 1e1, 1e2, 1e3, 1e4, 1e5},
    ytick={-0.6,-0.4,-0.2,0,0.2,0.4},
    xminorticks=false,
    yminorticks=false,
    axis lines=left,
    axis line style={-stealth, thick},
    tick align=outside,
    tick style={black, thick},
    label style={font=\small},
    tick label style={font=\small},
    grid=major,
    grid style={line width=0.1pt, draw=gray!12},
    major grid style={line width=0.2pt, draw=gray!18},
    enlargelimits=false,
    clip=true,
    axis on top=true,
    legend pos=north west,
    legend cell align=left,
    legend style={
        draw=gray!60,
        fill=white,
        fill opacity=0.92,
        text opacity=1,
        font=\scriptsize,
        inner sep=3pt,
        row sep=1pt,
        cells={anchor=west}
    }
]

\addplot[draw=none, fill=gray!7, forget plot]
coordinates {(500,-0.75) (500,0.6) (1e5,0.6) (1e5,-0.75)} \closedcycle;

\addplot[draw=none, fill=gray!7, forget plot]
coordinates {(500,-0.75) (1e5,-0.75) (1e5,0.0) (500,0.0)} \closedcycle;

\addplot[black, densely dashed, line width=0.8pt, forget plot]
coordinates {(1e0,0) (1e5,0)};

\addplot[black, densely dashed, line width=1.1pt, forget plot]
coordinates {(500,-0.75) (500,0.6)};

\addplot[initcol, densely dashed, line width=1.4pt]
table[x=omega, y=nu_0, col sep=comma]
{figures_data/cc_ifp.csv};
\addlegendentry{$\theta^{(0)}$}

\addplot[optcol, line width=1.7pt]
table[x=omega, y=nu_opt, col sep=comma]
{figures_data/cc_ifp.csv};
\addlegendentry{$\theta^{(34)}$}

\addplot[only marks, mark=*, mark size=2.5pt,
         mark options={fill=initcol, draw=white, line width=0.4pt}, forget plot]
coordinates {(500,-0.395)};
\addplot[only marks, mark=*, mark size=2.5pt,
         mark options={fill=optcol, draw=white, line width=0.4pt}, forget plot]
coordinates {(500,0.0)};


\node[anchor=west, font=\scriptsize, text=gray!80!black]
    at (axis cs:560,-0.68) {$\omega_1=500 \text{ rad/s}$};

\end{axis}
\end{tikzpicture}
\caption{Frequency-domain passivity profile for the initial parameter set~$\theta^{(0)}$ 
(orange, dotted) and the tuned set~$\theta^{(34)}$ (blue, solid), where
$\nu(\omega) = \lambda_{\min}\!\bigl((G(j\omega)+G^*(j\omega))/2\bigr)$.
The shaded gray region denotes the target frequency range~$\Omega = [500, +\infty)$\,rad/s.}
\label{fig:ifp_frequency}
\end{figure}

Figure~\ref{fig:ifp_frequency} shows the frequency-domain passivity profile before and after tuning. In the target band,
the \ac{ifp} index shifts from a negative value to approximately zero, pulling the zero-crossing of the curve to a lower frequency and thereby reducing the non-passive frequency range. This confirms the effectiveness of the passivity index optimization.
 
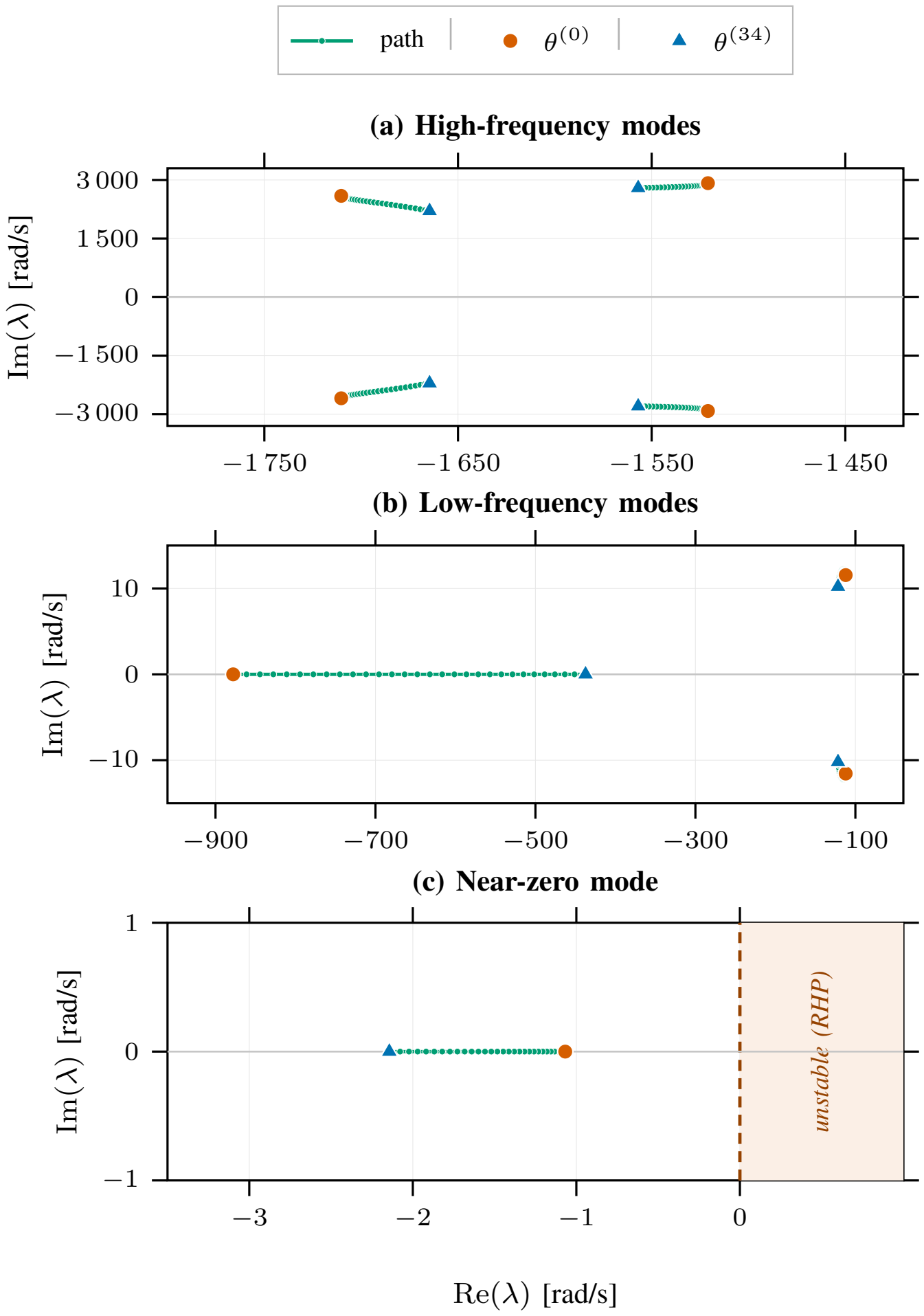
\begin{figure}[!t]
\centering
\definecolor{initcol}{RGB}{213,94,0}   
\definecolor{optcol}{RGB}{0,114,178}   
\definecolor{rhpcol}{RGB}{213,94,0}    
\definecolor{trajcol}{RGB}{0,158,115}  

\pgfplotsset{
    rl traj/.style={
        trajcol, line width=0.9pt, rounded corners=0.3pt,
        opacity=1, mark=*, mark size=0.9pt,
        mark options={fill=trajcol, draw=white, line width=0.2pt, solid},
    },
    theta0 marks/.style={
        color=initcol, only marks, mark=*, mark size=2.1pt,
        mark options={fill=initcol, draw=white, line width=0.4pt}
    },
    thetaopt marks/.style={
        color=optcol, only marks, mark=triangle*, mark size=2.7pt,
        mark options={fill=optcol, draw=white, line width=0.4pt}
    },
    sep bar/.style={
        draw=none,
        legend image code/.code={%
            \draw[gray!60, line width=0.5pt] (0.2cm,-0.55ex) -- (0.2cm,1.4ex);%
        },
    },
    rl axis/.style={
        width=0.80\columnwidth,
        height=0.28\columnwidth,
        scale only axis=true,
        anchor=south west,
        axis lines=box,
        axis line style={black, semithick},
        xlabel={$\operatorname{Re}(\lambda)$~[rad/s]},
        ylabel={$\operatorname{Im}(\lambda)$~[rad/s]},
        xlabel style={font=\footnotesize, yshift=1pt},
        ylabel style={font=\footnotesize, yshift=-2pt},
        tick align=outside,
        tick style={black, semithick},
        tick label style={font=\scriptsize},
        grid=major,
        major grid style={line width=0.2pt, draw=gray!18},
        enlargelimits=false,
        clip=true,
        axis on top=false,
        title style={font=\footnotesize\bfseries, yshift=-2pt},
    }
}

\begin{tikzpicture}

\begin{axis}[
    rl axis,
    name=panelA,
    xlabel={},
    at={(0,0.82\columnwidth)},
    xmin=-1800, xmax=-1420,
    ymin=-3300, ymax=3300,
    xtick={-1750,-1650,-1550,-1450},
    ytick={-3000,-1500,0,1500,3000},
    xticklabel style={/pgf/number format/.cd, fixed, set thousands separator={\,}},
    yticklabel style={/pgf/number format/.cd, fixed, set thousands separator={\,}},
    title={(a)~High-frequency modes},
    legend to name=cceiglegend,
    legend style={
        draw=gray!55,
        fill=white,
        font=\scriptsize,
        inner sep=3pt,
        column sep=5pt,
        cells={anchor=west},
    },
    legend columns=-1,
]

\draw[gray!45, line width=0.5pt] (axis cs:-1800,0) -- (axis cs:-1420,0);

\addplot[rl traj] table[x=real_1, y=imag_1, col sep=comma] {figures_data/cc_root_locus.csv};
\addlegendentry{path}
\addlegendimage{sep bar}
\addlegendentry{}
\addlegendimage{theta0 marks}
\addlegendentry{$\theta^{(0)}$}
\addlegendimage{sep bar}
\addlegendentry{}
\addlegendimage{thetaopt marks}
\addlegendentry{$\theta^{(34)}$}
\addplot[rl traj, forget plot] table[x=real_2, y=imag_2, col sep=comma] {figures_data/cc_root_locus.csv};
\addplot[rl traj, forget plot] table[x=real_3, y=imag_3, col sep=comma] {figures_data/cc_root_locus.csv};
\addplot[rl traj, forget plot] table[x=real_4, y=imag_4, col sep=comma] {figures_data/cc_root_locus.csv};

\addplot[theta0 marks, forget plot,
    restrict expr to domain={\thisrow{real}}{-1800:-1420}]
table[x=real, y=imag, col sep=comma] {figures_data/cc_eig_0.csv};

\addplot[thetaopt marks, forget plot,
    restrict expr to domain={\thisrow{real}}{-1800:-1420}]
table[x=real, y=imag, col sep=comma] {figures_data/cc_eig_opt.csv};

\end{axis}

\begin{axis}[
    rl axis,
    xlabel={},
    at={(0,0.41\columnwidth)},
    xmin=-960, xmax=-40,
    ymin=-15, ymax=15,
    xtick={-900,-700,-500,-300,-100},
    ytick={-10,0,10},
    title={(b)~Low-frequency modes},
]

\draw[gray!45, line width=0.5pt] (axis cs:-960,0) -- (axis cs:-40,0);

\addplot[rl traj, forget plot] table[x=real_5, y=imag_5, col sep=comma] {figures_data/cc_root_locus.csv};
\addplot[rl traj, forget plot] table[x=real_6, y=imag_6, col sep=comma] {figures_data/cc_root_locus.csv};
\addplot[rl traj, forget plot] table[x=real_7, y=imag_7, col sep=comma] {figures_data/cc_root_locus.csv};

\addplot[theta0 marks,
    restrict expr to domain={\thisrow{real}}{-960:-40}]
table[x=real, y=imag, col sep=comma] {figures_data/cc_eig_0.csv};

\addplot[thetaopt marks,
    restrict expr to domain={\thisrow{real}}{-960:-40}]
table[x=real, y=imag, col sep=comma] {figures_data/cc_eig_opt.csv};

\end{axis}

\begin{axis}[
    rl axis,
    at={(0,0)},
    xmin=-3.5, xmax=1,
    ymin=-1, ymax=1,
    xtick={-3,-2,-1},
    extra x ticks={0},
    extra x tick labels={$0$},
    ytick={-1,0,1},
    title={(c)~Near-zero mode},
]

\addplot[draw=none, fill=rhpcol!10, forget plot]
coordinates {(0,-1) (1,-1) (1,1) (0,1)} \closedcycle;
\draw[rhpcol!70!black, densely dashed, line width=0.9pt]
    (axis cs:0,-1) -- (axis cs:0,1);
\node[rotate=90, anchor=center, font=\scriptsize\itshape, text=rhpcol!70!black]
    at (axis cs:0.5,0) {unstable (RHP)};

\draw[gray!45, line width=0.5pt] (axis cs:-3.5,0) -- (axis cs:1,0);

\addplot[rl traj, forget plot] table[x=real_8, y=imag_8, col sep=comma] {figures_data/cc_root_locus.csv};

\addplot[theta0 marks,
    restrict expr to domain={\thisrow{real}}{-3.5:1}]
table[x=real, y=imag, col sep=comma] {figures_data/cc_eig_0.csv};

\addplot[thetaopt marks,
    restrict expr to domain={\thisrow{real}}{-3.5:1}]
table[x=real, y=imag, col sep=comma] {figures_data/cc_eig_opt.csv};

\end{axis}

\node[anchor=south, yshift=4pt]
    at (panelA.north |- current bounding box.north)
    {\pgfplotslegendfromname{cceiglegend}};

\end{tikzpicture}
\caption{Eigenvalues of the linearized \ac{cc} converter system matrix~$A$ 
for the initial parameter set~$\theta^{(0)}$ ($\bullet$, orange) 
and the tuned set~$\theta^{(34)}$ ($\blacktriangle$, blue).
Green dotted lines show the transition path during optimization iterations. All poles remain in the open left-half plane, 
confirming preserved stability.}
\label{fig:eigenvalues}
\end{figure}

Figure~\ref{fig:eigenvalues} shows the eigenvalues of $A(\theta)$
before and after optimization. All eight poles remain in the open
left-half plane, consistent with the Lyapunov stability constraint
imposed in~\eqref{eq:stab_lmi}.
It is worth noting that after the optimization, the damping of the eigenvalues has
generally increased, which is in line with the interpretation of passivity as damping~\cite{Harnefors2016Passivity}.

\subsection{Case~2: Multi-Converter Microgrid}\label{sec:case_mc}
 
\subsubsection{System Description}
 
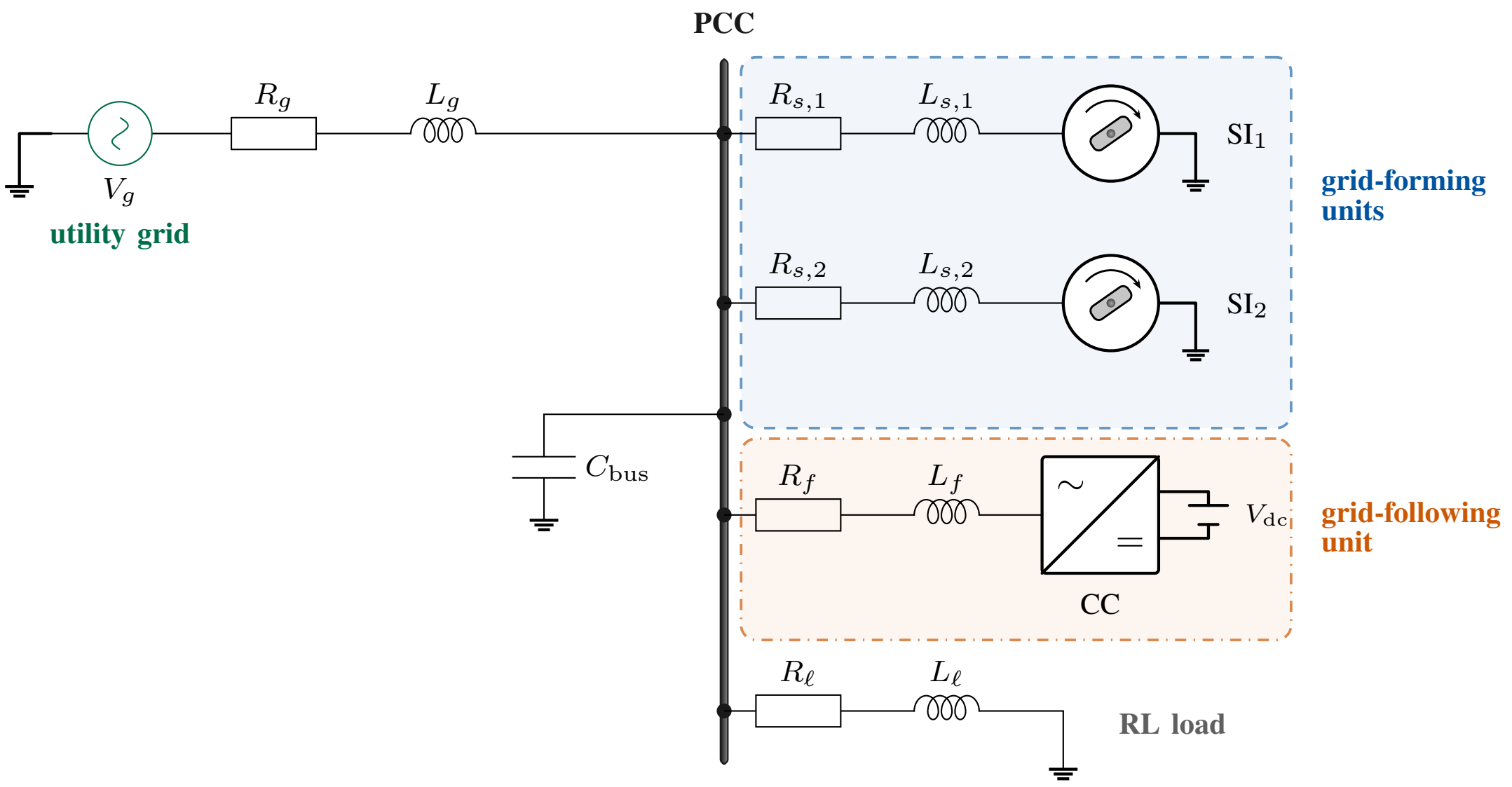
\begin{figure*}[!t]
\centering
\begingroup
\ctikzset{bipoles/thickness=1, inductors/coils=4, inductor=cute,
          bipoles/length=1.0cm}%
\begin{tikzpicture}[
  x=1cm,y=1cm,
  font=\footnotesize,
  wire/.style={line width=0.8pt, black, line cap=round, line join=round},
  tap/.style={circle, fill=busblack, inner sep=0pt, minimum size=4pt},
  cvlabel/.style={font=\footnotesize, align=center},
]
 
\newcommand{\vsc}[1]{%
  \begin{scope}[shift={(#1)}]
    \draw[wire, fill=white, rounded corners=1pt] (-0.55,-0.55) rectangle (0.55,0.55);
    \draw[wire] (-0.55,-0.55) -- (0.55,0.55);
    \node[font=\normalsize] at (-0.28,0.26) {$\sim$};
    \node[font=\normalsize] at (0.28,-0.28) {$=$};
    \draw[wire] (0.55,0.22) -- (1.02,0.22) -- (1.02,0.09);
    \draw[wire] (0.55,-0.22) -- (1.02,-0.22) -- (1.02,-0.09);
    \draw[wire] (0.84,0.09) -- (1.20,0.09);    
    \draw[wire] (0.93,-0.09) -- (1.11,-0.09);  
    \node[font=\scriptsize, anchor=west] at (1.24,0) {$V_{\mathrm{dc}}$};
  \end{scope}
}
 
\newcommand{\vsm}[1]{%
  \begin{scope}[shift={(#1)}]
    \draw[wire, fill=white] (0,0) circle (0.45);
    \begin{scope}[rotate=35]
      \draw[line width=0.55pt, fill=gray!45, rounded corners=1.6pt]
        (-0.21,-0.075) rectangle (0.21,0.075);
    \end{scope}
    \shade[inner color=black!35, outer color=black!70] (0,0) circle (0.042);   
    \draw[line width=0.55pt, -{Stealth[length=2.6pt]}]
      (140:0.31) arc[start angle=140, end angle=25, radius=0.31];
    \draw[wire] (0.45,0) -- (0.80,0) -- (0.80,-0.45);
    \node[tlground, scale=0.85] at (0.80,-0.45) {};
  \end{scope}
}
 
\def\busx{0}
\def\ySVa{1.05}
\def\ySVb{-0.55}
\def\yCC{-2.55}
\def\yLD{-4.40}
\def\yG{1.05}
\def\yC{-1.60}
\def\bustop{1.75}
\def\busbot{-4.90}
 
\shade[left color=busblack, right color=black!45, draw=busblack, line width=0.4pt,
       rounded corners=1.3pt] (\busx-0.035,\busbot) rectangle (\busx+0.035,\bustop);
\node[busblack, font=\footnotesize\bfseries, anchor=south] at (\busx,\bustop+0.12) {PCC};
 
\coordinate (gt) at (\busx,\yG); \node[tap] at (gt) {};
\draw[wire] (-6.35,\yG) -- (-6.65,\yG) -- (-6.65,\yG-0.50);
\node[tlground, scale=0.9] at (-6.65,\yG-0.50) {};
\draw (-6.35,\yG) to[sV, l_=$V_g$, color=gridgreen] (-5.05,\yG)
      to[R, l=$R_g$] (-3.45,\yG)
      to[L, l=$L_g$] (-1.85,\yG) -- (gt);
\node[gridgreen, font=\footnotesize\bfseries, anchor=north] at (-5.70,\yG-0.72) {utility grid};

\coordinate (cbt) at (\busx,\yC); \node[tap] at (cbt) {};
\draw (cbt) -- (-1.70,\yC) to[C, l=$C_{\mathrm{bus}}$] ++(0,-1.00)
      node[tlground, scale=0.9]{};
 
\coordinate (t1) at (\busx,\ySVa); \node[tap] at (t1) {};
\draw (t1) to[R, l=$R_{s,1}$] (1.4,\ySVa) to[L, l=$L_{s,1}$] (2.8,\ySVa) -- (3.20,\ySVa);
\vsm{(3.65,\ySVa)}
\node[cvlabel, anchor=west] at (4.62,\ySVa-0.02) {SI$_1$};
 
\coordinate (t2) at (\busx,\ySVb); \node[tap] at (t2) {};
\draw (t2) to[R, l=$R_{s,2}$] (1.4,\ySVb) to[L, l=$L_{s,2}$] (2.8,\ySVb) -- (3.20,\ySVb);
\vsm{(3.65,\ySVb)}
\node[cvlabel, anchor=west] at (4.62,\ySVb-0.02) {SI$_2$};
 
\coordinate (t3) at (\busx,\yCC); \node[tap] at (t3) {};
\draw (t3) to[R, l=$R_{f}$] (1.4,\yCC) to[L, l=$L_{f}$] (2.8,\yCC) -- (3.0,\yCC);
\vsc{(3.55,\yCC)}
\node[cvlabel, anchor=north] at (3.55,\yCC-0.62) {CC};
 
\coordinate (t4) at (\busx,\yLD); \node[tap] at (t4) {};
\draw (t4) to[R, l=$R_{\ell}$] (1.4,\yLD) to[L, l=$L_{\ell}$] (2.8,\yLD)
      -- (3.2,\yLD) -- (3.2,\yLD-0.55) node[tlground, scale=0.9]{};
\node[loadgray, font=\footnotesize\bfseries, anchor=west] at (3.60,\yLD-0.12) {RL load};
 
\begin{scope}[on background layer]
  \node[draw=gfmblue!60, dashed, line width=0.7pt, rounded corners=4pt,
        fill=gfmblue!5, inner sep=0pt,
        fit={(0.16,\ySVa+0.72) (5.35,\ySVb-1.18)}] (gfmbox) {};
  \node[draw=gflorange!70, dash dot, line width=0.7pt, rounded corners=4pt,
        fill=gflorange!6, inner sep=0pt,
        fit={(0.16,\yCC+0.72) (5.35,\yCC-1.18)}] (gflbox) {};
\end{scope}
\node[gfmblue, font=\footnotesize\bfseries, anchor=west, align=left]
  at ($(gfmbox.east)+(0.15,0.45)$) {grid-forming\\[-1pt]units};
\node[gflorange, font=\footnotesize\bfseries, anchor=west, align=left]
  at ($(gflbox.east)+(0.15,0.12)$) {grid-following\\[-1pt]unit};
 
\end{tikzpicture}%
\endgroup
\caption{Topology of the multi-converter microgrid in  the $dq$ reference frame: two synchronverters (SI), 
 one grid-following current controlled (CC) converter, and a $RL$ load connected to a common bus with shunt capacitor~$C_{\mathrm{bus}}$.}
\label{fig:mc_topology}
\end{figure*}

The second case study considers an AC microgrid modeled in the $dq$ reference frame; see Fig.~\ref{fig:mc_topology}. An ideal grid voltage source is connected to the \ac{pcc} through a series $RL$ line. At the \ac{pcc}, two synchronverters~\cite{Zhong2011Synchronverters}, one grid-following current-controlled converter, a shunt bus capacitor,
and a passive $RL$ load are connected in parallel.
 
Table~\ref{tab:mc_plant_params} lists the system parameters. Linearization around the nominal dispatch produces
a state-space realization \(
(A,B,C,D) \in
\R^{24\times 24}
\times\R^{24\times 2}
\times\R^{2\times 24}
\times\R^{2\times 2}
\) with inputs \((V_d, V_q)\), the $dq$ components of the utility grid voltage, and outputs \((i_d, i_q)\), the $dq$ components of the current flowing from the grid into the \ac{pcc} through the coupling line~$(R_g, L_g)$. This input--output pair defines the admittance port model of the microgrid, at which we evaluate passivity.

\begin{table}[!t]
\centering
\caption{System parameters for the multi-converter microgrid.}
\label{tab:mc_plant_params}
\scriptsize
\begin{tabular}{lll}
\toprule
Symbol & Description & Value \\
\midrule
\multicolumn{3}{l}{\emph{Grid and coupling line}} \\
$V_d$                  & $d$-axis grid voltage      & $170$\,V \\
$V_q$                  & $q$-axis grid voltage      & $0$\,V \\
$R_g$                  & Line resistance            & $0.5\,\Omega$ \\
$L_g$                  & Line inductance            & $3$\,mH \\
$C_{\mathrm{bus}}$     & Bus capacitor              & $20\,\mu$F \\
$\omega_g$             & Grid frequency             & $2\pi\cdot50$\,rad/s \\
\midrule
\multicolumn{3}{l}{\emph{Synchronverter 1}} \\
$P_{\mathrm{set},1}$   & Active power setpoint      & $500$\,W \\
$Q_{\mathrm{set},1}$   & Reactive power setpoint    & $300$\,VAr \\
$R_{s,1}$              & Filter resistance          & $0.7\,\Omega$ \\
$L_{s,1}$              & Filter inductance          & $2$\,mH \\
\midrule
\multicolumn{3}{l}{\emph{Synchronverter 2}} \\
$P_{\mathrm{set},2}$   & Active power setpoint      & $300$\,W \\
$Q_{\mathrm{set},2}$   & Reactive power setpoint    & $150$\,VAr \\
$R_{s,2}$              & Filter resistance          & $0.8\,\Omega$ \\
$L_{s,2}$              & Filter inductance          & $2.5$\,mH \\
\midrule
\multicolumn{3}{l}{\emph{Grid-following converter}} \\
$V_{\mathrm{dc}}$      & DC-link voltage            & $600$\,V \\
$L_f$                    & Filter inductance          & $1$\,mH \\
$R_f$                    & Filter resistance          & $10$\,m$\Omega$ \\
$f_s$                  & Switching frequency        & $5$\,kHz \\
$T_d$                  & PWM delay                  & $300\,\mu$s \\
$i_d^{\mathrm{ref}}$   & $d$-axis current reference & $3.0$\,A \\
$i_q^{\mathrm{ref}}$   & $q$-axis current reference & $0$\,A \\
\midrule
\multicolumn{3}{l}{\emph{Passive load}} \\
$R_{\ell}$    & Load resistance            & $30\,\Omega$ \\
$L_{\ell}$    & Load inductance            & $10$\,mH \\
\bottomrule
\end{tabular}
\end{table}

\subsubsection{Tunable Parameters and Bounds}
The parameter vector contains fourteen tunable parameters:
\[
\begin{aligned}
  \theta = \bigl[
  &J_1,\; D_{p,1},\; K_1,\; D_{q,1},\;
  J_2,\; D_{p,2},\; K_2,\; D_{q,2}, \\
  &k_p^{d},\; k_i^{d},\; k_p^{q},\; k_i^{q},\;
  k_p^{\mathrm{PLL}},\; k_i^{\mathrm{PLL}}
  \bigr]^\top .
\end{aligned}
\]
The first eight entries are the virtual inertia, active damping,
reactive control gain, and reactive damping of the two
synchronverters. The remaining six entries are the $dq$-axis
\ac{cc} \ac{pi} gains and \ac{pll} \ac{pi} gains of the grid-following
converter. The vector is initialized at
\[
\begin{aligned}
  \theta^{(0)} = \bigl[
  &0.0380,\; 3.799,\; 1852,\; 589.4,\;
  0.0228,\; 2.279,\; 925.8,\; \\
  &294.7,\;0.0105,\; 1.152,\; 0.0105,\; 1.152,\;
  3.0,\; 3.2
  \bigr]^\top .
\end{aligned}
\]
The synchronverter parameters are initialized according to the design
guidelines in~\cite{Zhong2011Synchronverters}, scaled to the
rated power of the microgrid.
The objective is to maximize the \ac{ifp} index over
\[
    \Omega=[\omega_1,\omega_2]=[35,50]\ \mathrm{rad/s}.
\]
We select this range (see the gray shaded region in Fig.~\ref{fig:ifp_mc}) to test whether the method can further increase the already positive maximum 
of the nominal frequency-dependent \ac{ifp} profile.
This sub-synchronous interval (approximately 5--8\,Hz) targets the range
where finite grid impedance, bus capacitance, and synchronverter swing
dynamics produce dominant converter-network
interactions~\cite{BNetzA2025,ENTSOE_TGGFC2025}.
Tables~\ref{tab:mc_params} and~\ref{tab:hyperparams_mc} list the
bounds and hyperparameters.

\begin{table}[!t]
\centering
\caption{Box constraints on the tunable parameters in Case~2.}
\label{tab:mc_params}
\scriptsize
\begin{tabular}{lc}
\toprule
Parameter & $\Theta_j$ \\
\midrule
$J_1$                    & $[0.0190,\; 0.0570]$ \\
$D_{p,1}$                & $[1.899,\; 5.698]$ \\
$K_1$                    & $[925.8,\; 2777.5]$ \\
$D_{q,1}$                & $[294.7,\; 884.1]$ \\
$J_2$                    & $[0.0114,\; 0.0342]$ \\
$D_{p,2}$                & $[1.140,\; 3.419]$ \\
$K_2$                    & $[462.9,\; 1388.7]$ \\
$D_{q,2}$                & $[147.4,\; 442.1]$ \\
$k_p^{d}$                & $[0.0053,\; 0.0158]$ \\
$k_i^{d}$                & $[0.576,\; 1.728]$ \\
$k_p^{q}$                & $[0.0053,\; 0.0158]$ \\
$k_i^{q}$                & $[0.576,\; 1.728]$ \\
$k_p^{\mathrm{PLL}}$     & $[1.5,\; 4.5]$ \\
$k_i^{\mathrm{PLL}}$     & $[1.6,\; 4.8]$ \\
\bottomrule
\end{tabular}
\end{table}
 
\begin{table}[!t]
\centering
\caption{Algorithm hyperparameters in Case~2.}
\label{tab:hyperparams_mc}
\begin{tabular}{lll}
\toprule
Symbol & Description & Value \\
\midrule
$\eta_0$       & Initial step size    & $5.0$ \\
$\beta$        & Backtracking factor  & $0.3$ \\
$c$            & Armijo constant      & $0.1$ \\
               & Maximum backtracks   & $4$ \\
$\varepsilon_g$ & Stationarity tolerance & $10^{-2}$ \\
\bottomrule
\end{tabular}
\end{table}

\subsubsection{Results}\label{sec:mc_results}
 
\begin{figure}[!t]
\centering
\begin{tikzpicture}
\definecolor{optcol}{RGB}{0,114,178}    
\definecolor{threshcol}{RGB}{118,42,131} 

\begin{axis}[
    width=\columnwidth,
    height=0.72\columnwidth,
    xlabel={Iteration $k$},
    ylabel={IFP index $\nu(\theta^{(k)})$},
    xmin=0, xmax=13,
    ymin=0.25, ymax=0.33,
    xtick={0,3,6,9,13},
    ytick={0.26,0.28,0.30, 0.32},
    grid=major,
    grid style={line width=0.1pt, draw=gray!12},
    major grid style={line width=0.2pt, draw=gray!18},
    axis lines=left,
    axis line style={-stealth, thick},
    tick align=outside,
    tick style={black, thick},
    label style={font=\small},
    tick label style={font=\small},
    enlargelimits=false,
    clip=true,
    axis on top=true,
    legend pos=south east,
    legend cell align=left,
    legend style={
        draw=gray!60,
        fill=white,
        fill opacity=0.92,
        text opacity=1,
        font=\scriptsize,
        inner sep=3pt,
        row sep=1pt,
        cells={anchor=west}
    }
]


\addplot[optcol, line width=1.4pt, mark=*, mark size=1.8pt,
         mark options={fill=optcol, draw=white, line width=0.3pt}]
table[x=iteration, y=nu, col sep=comma]
{figures_data/mc_obj.csv};
\addlegendentry{$\nu(\theta^{(k)})$}

\end{axis}
\end{tikzpicture}
\caption{Minimum \ac{ifp} index~$\nu$
over the frequency range of interest of the multi-converter microgrid across the optimization iterations.}
\label{fig:obj_mc}
\end{figure}
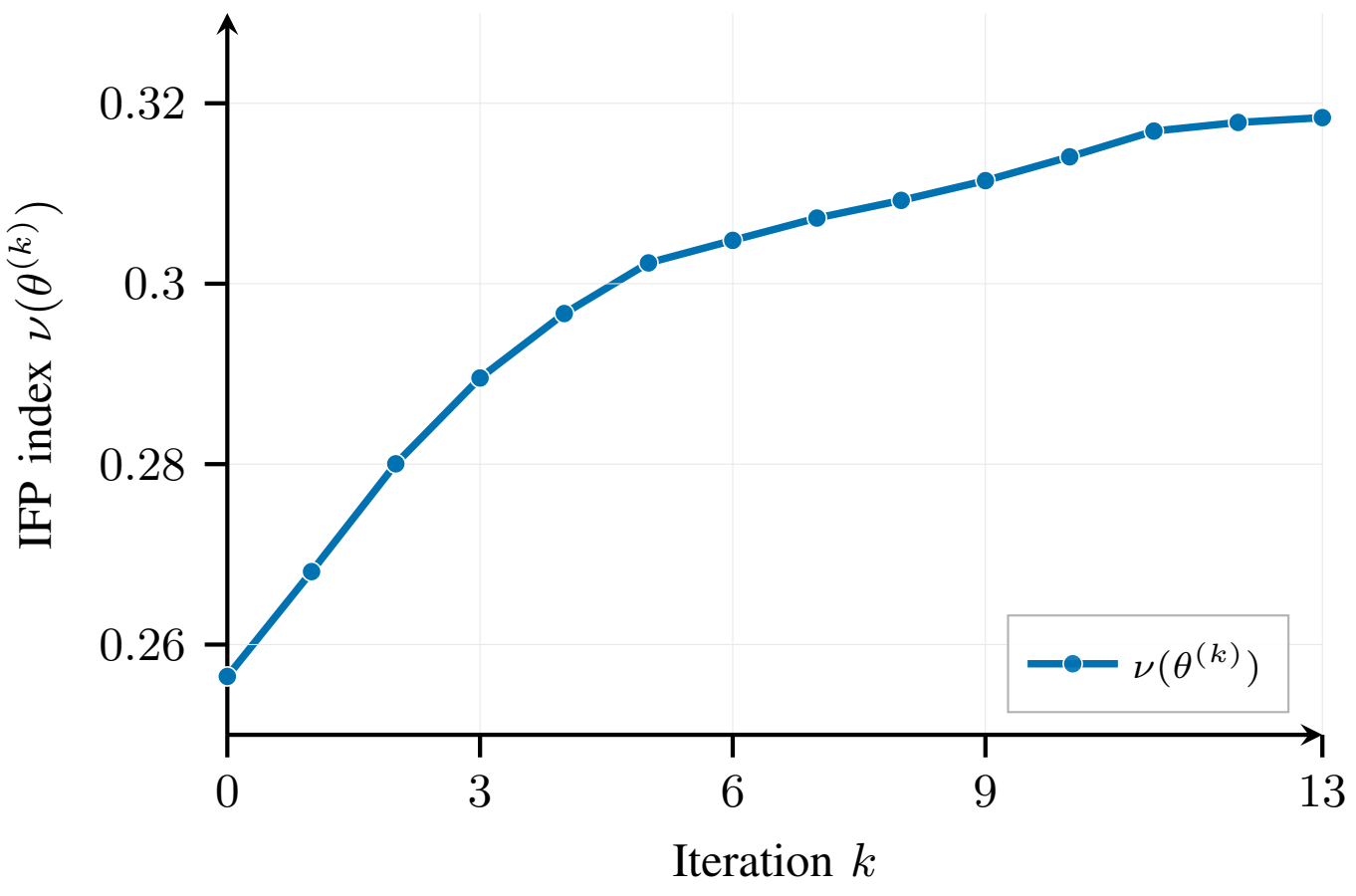

Figure~\ref{fig:obj_mc} shows the minimum \ac{ifp} index over the frequency
range of interest across the optimization iterations. The algorithm converges
in 13 iterations, increasing the index monotonically
from $\nu(\theta^{(0)})=0.2565$ to $\nu(\theta^{(13)})=0.3184$
(approximately $24.1\%$ improvement in the target band).
 
The system is already passive over $\Omega$ at the nominal values $\theta^{(0)}$. Therefore, the
increase in $\nu$ represents a damping and robustness improvement in that frequency range, not a
correction of an initial passivity violation. 
 
Table~\ref{tab:mc_tuning_results} reports the optimized parameter values. The tuning acts mainly on the synchronverter parameters. For synchronverter~1, $J_1$, $K_1$, and $D_{q,1}$ reach their upper bounds, while $D_{p,1}$ saturates its lower bound. For synchronverter~2, $J_2$ and $D_{q,2}$ reach their upper bounds, while $D_{p,2}$ and $K_2$ decrease by $43.2\%$ and
$15.3\%$, respectively. In contrast, all grid-following current-control and \ac{pll} gains change by less than $4\%$.
 
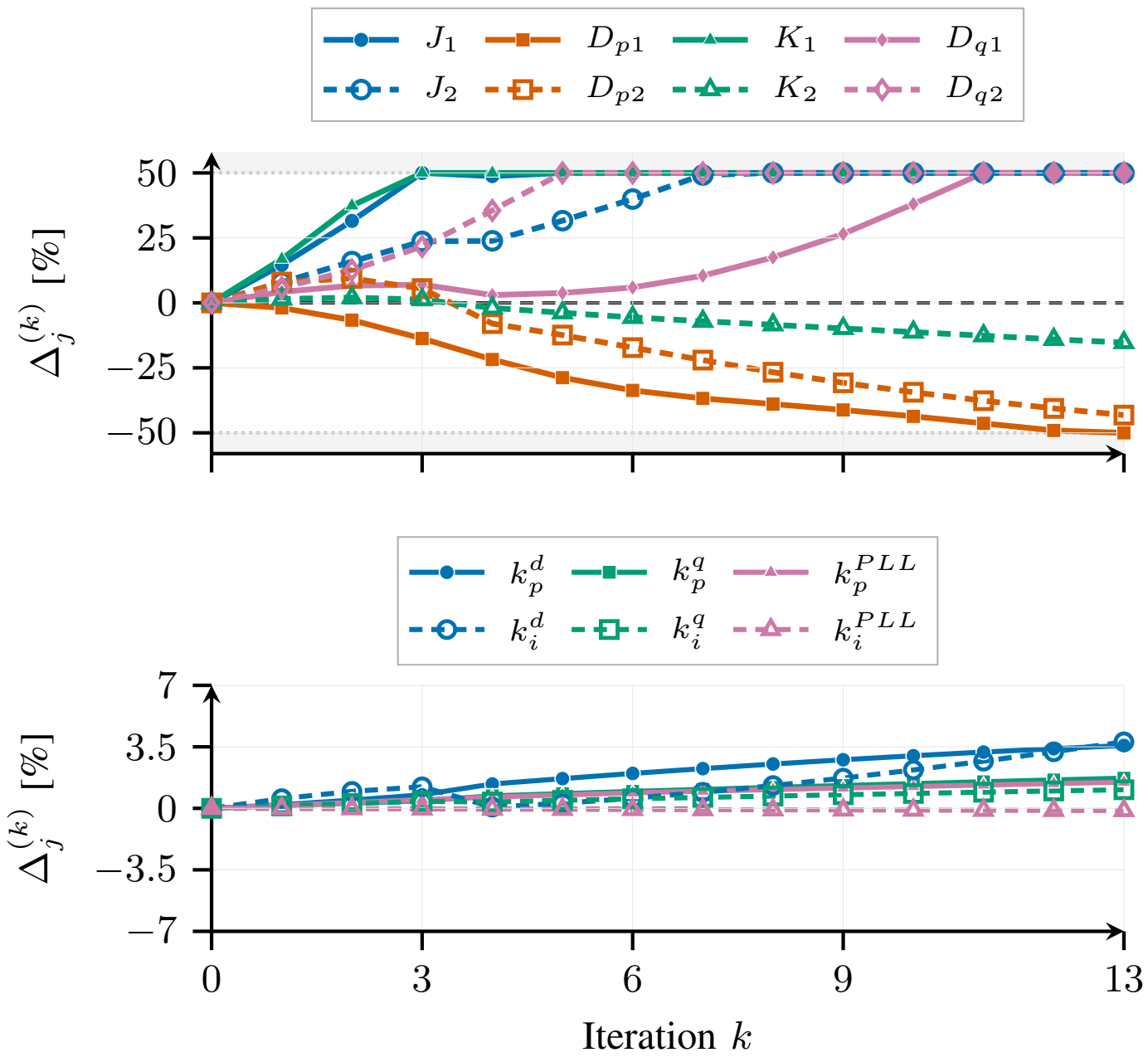
\begin{figure}[!t]
\centering
\begin{tikzpicture}
\definecolor{cJ}{RGB}{0,114,178}   
\definecolor{cDp}{RGB}{213,94,0}   
\definecolor{cK}{RGB}{0,158,115}   
\definecolor{cDq}{RGB}{204,121,167}
\definecolor{cE}{RGB}{230,159,0}   
\definecolor{cF}{RGB}{86,180,233}  

\pgfplotsset{
    filledmark/.style={mark options={fill=#1, draw=white, line width=0.3pt, solid}},
    openmark/.style={mark options={fill=white, draw=#1, line width=0.9pt, solid}},
}

\begin{axis}[
    width=\columnwidth,
    height=0.45\columnwidth,
    at={(0,0.43\columnwidth)},
    anchor=south west,
    xlabel={},
    ylabel={$\Delta_j^{(k)}$ [\%]},
    xmin=0,
    xmax=13,
    ymin=-58,
    ymax=58,
    xtick={0,3,6,9,13},
    xticklabels={},
    ytick={-50,-25,0,25,50},
    grid=major,
    axis lines=left,
    axis line style={-stealth, thick},
    tick align=outside,
    tick style={black, thick},
    label style={font=\small},
    tick label style={font=\small},
    grid style={line width=0.1pt, draw=gray!12},
    major grid style={line width=0.2pt, draw=gray!18},
    enlargelimits=false,
    clip=true,
    axis on top=true,
    legend columns=4,
    legend style={
        at={(0.5,1.10)},
        anchor=south,
        draw=gray!60,
        fill=white,
        font=\scriptsize,
        inner sep=2pt,
        column sep=3pt,
        row sep=1pt
    }
]

\addplot[draw=none, fill=gray!9, forget plot]
coordinates {(0,50) (18,50) (18,58) (0,58)} \closedcycle;
\addplot[draw=none, fill=gray!9, forget plot]
coordinates {(0,-58) (18,-58) (18,-50) (0,-50)} \closedcycle;
\addplot[gray!55, densely dotted, line width=0.7pt, forget plot]
coordinates {(0,50) (18,50)};
\addplot[gray!55, densely dotted, line width=0.7pt, forget plot]
coordinates {(0,-50) (18,-50)};

\addplot[black, densely dashed, line width=0.7pt, forget plot]
coordinates {(0,0) (18,0)};

\addplot[color=cJ, line width=1.3pt, solid, mark=*, mark size=1.7pt,
         mark options={fill=cJ, draw=white, line width=0.2pt}]
table[x=iteration, y=J_1_pct, col sep=comma]
{figures_data/mc_par.csv};
\addlegendentry{$J_1$}

\addplot[color=cDp, line width=1.3pt, solid, mark=square*, mark size=1.6pt,
         mark options={fill=cDp, draw=white, line width=0.2pt}]
table[x=iteration, y=Dp_1_pct, col sep=comma]
{figures_data/mc_par.csv};
\addlegendentry{$D_{p1}$}

\addplot[color=cK, line width=1.3pt, solid, mark=triangle*, mark size=1.9pt,
         mark options={fill=cK, draw=white, line width=0.2pt}]
table[x=iteration, y=K_1_pct, col sep=comma]
{figures_data/mc_par.csv};
\addlegendentry{$K_1$}

\addplot[color=cDq, line width=1.3pt, solid, mark=diamond*, mark size=1.8pt,
         mark options={fill=cDq, draw=white, line width=0.2pt}]
table[x=iteration, y=Dq_1_pct, col sep=comma]
{figures_data/mc_par.csv};
\addlegendentry{$D_{q1}$}

\addplot[color=cJ, line width=1.3pt, densely dashed, mark=o, mark size=2.0pt,
         openmark=cJ]
table[x=iteration, y=J_2_pct, col sep=comma]
{figures_data/mc_par.csv};
\addlegendentry{$J_2$}

\addplot[color=cDp, line width=1.3pt, densely dashed, mark=square, mark size=1.9pt,
         openmark=cDp]
table[x=iteration, y=Dp_2_pct, col sep=comma]
{figures_data/mc_par.csv};
\addlegendentry{$D_{p2}$}

\addplot[color=cK, line width=1.3pt, densely dashed, mark=triangle, mark size=2.1pt,
         openmark=cK]
table[x=iteration, y=K_2_pct, col sep=comma]
{figures_data/mc_par.csv};
\addlegendentry{$K_2$}

\addplot[color=cDq, line width=1.3pt, densely dashed, mark=diamond, mark size=2.1pt,
         openmark=cDq]
table[x=iteration, y=Dq_2_pct, col sep=comma]
{figures_data/mc_par.csv};
\addlegendentry{$D_{q2}$}

\end{axis}

\begin{axis}[
    width=\columnwidth,
    height=0.40\columnwidth,
    at={(0,0)},
    anchor=south west,
    xlabel={Iteration $k$},
    ylabel={$\Delta_j^{(k)}$ [\%]},
    xmin=0,
    xmax=13,
    ymin=-7.00,
    ymax=7.00,
    xtick={0,3,6,9,13},
    ytick={-7,-3.5,0,3.5,7},
    grid=major,
    axis lines=left,
    axis line style={-stealth, thick},
    tick align=outside,
    tick style={black, thick},
    label style={font=\small},
    tick label style={font=\small},
    grid style={line width=0.1pt, draw=gray!12},
    major grid style={line width=0.2pt, draw=gray!18},
    enlargelimits=false,
    clip=true,
    axis on top=true,
    legend columns=3,
    legend style={
        at={(0.5,1.08)},
        anchor=south,
        draw=gray!60,
        fill=white,
        font=\scriptsize,
        inner sep=2pt,
        column sep=3pt,
        row sep=1pt
    }
]

\addplot[black, densely dashed, line width=0.7pt, forget plot]
coordinates {(0,0) (7,0)};

\addplot[color=cJ, line width=1.15pt, solid, mark=*, mark size=1.7pt,
         filledmark=cJ]
table[x=iteration, y=kp_CC_d_pct, col sep=comma]
{figures_data/mc_par.csv};
\addlegendentry{$k_p^d$}

\addplot[color=cK, line width=1.15pt, solid, mark=square*, mark size=1.6pt,
         filledmark=cK]
table[x=iteration, y=kp_CC_q_pct, col sep=comma]
{figures_data/mc_par.csv};
\addlegendentry{$k_p^q$}

\addplot[color=cDq, line width=1.15pt, solid, mark=triangle*, mark size=1.9pt,
         filledmark=cDq]
table[x=iteration, y=kp_PLL_pct, col sep=comma]
{figures_data/mc_par.csv};
\addlegendentry{$k_p^{PLL}$}

\addplot[color=cJ, line width=1.15pt, densely dashed, mark=o, mark size=1.9pt,
         openmark=cJ]
table[x=iteration, y=ki_CC_d_pct, col sep=comma]
{figures_data/mc_par.csv};
\addlegendentry{$k_i^d$}

\addplot[color=cK, line width=1.15pt, densely dashed, mark=square, mark size=1.9pt,
         openmark=cK]
table[x=iteration, y=ki_CC_q_pct, col sep=comma]
{figures_data/mc_par.csv};
\addlegendentry{$k_i^q$}

\addplot[color=cDq, line width=1.15pt, densely dashed, mark=triangle, mark size=2.1pt,
         openmark=cDq]
table[x=iteration, y=ki_PLL_pct, col sep=comma]
{figures_data/mc_par.csv};
\addlegendentry{$k_i^{PLL}$}

\end{axis}
\end{tikzpicture}
\caption{Relative change~$\Delta_j^{(k)}$~[\%] in microgrid parameters at each optimization iteration.}
\label{fig:par_mc}
\end{figure}

Figure~\ref{fig:par_mc} displays the relative change~$\Delta_j^{(k)}$ of the synchronverter (top) and grid-following converter (bottom) parameters across iterations. The dominant tuning directions are associated with the grid-forming units, whose inertia and damping parameters directly shape the passivity characteristics in the targeted frequency range. The parameters of the grid-following converter have less influence on the passivity at that frequency range and thus do not change.
\begin{table}[!t]
\centering
\caption{Tuning results: initial versus tuned gains in Case~2.
Parameters marked with $\dagger$ saturate a bound.}
\label{tab:mc_tuning_results}
\scriptsize
\begin{tabular}{lccc}
\toprule
Parameter & $\theta_j^{(0)}$ & $\theta_j^{(13)}$ & Change \\
\midrule
$J_1^{\dagger}$             & $0.0380$  & $0.0570$  & $+50.0\%$ \\
$D_{p,1}^{\dagger}$         & $3.799$   & $1.899$   & $-50.0\%$ \\
$K_1^{\dagger}$             & $1852$    & $2777.5$  & $+50.0\%$ \\
$D_{q,1}^{\dagger}$         & $589.4$   & $884.1$   & $+50.0\%$ \\
$J_2^{\dagger}$             & $0.0228$  & $0.0342$  & $+50.0\%$ \\
$D_{p,2}$                   & $2.279$   & $1.295$   & $-43.2\%$ \\
$K_2$                       & $925.8$   & $784.5$   & $-15.3\%$ \\
$D_{q,2}^{\dagger}$         & $294.7$   & $442.1$   & $+50.0\%$ \\
$k_p^{d}$                   & $0.0105$  & $0.01088$ & $+3.6\%$ \\
$k_i^{d}$                   & $1.1519$   & $1.195$   & $+3.8\%$ \\
$k_p^{q}$                   & $0.0105$  & $0.01068$ & $+1.7\%$ \\
$k_i^{q}$                   & $1.152$   & $1.164$   & $+1.1\%$ \\
$k_p^{\mathrm{PLL}}$        & $3.0$     & $3.045$   & $+1.5\%$ \\
$k_i^{\mathrm{PLL}}$        & $3.2$     & $3.196$   & $-0.1\%$ \\
\bottomrule
\end{tabular}
\end{table}

\begin{figure}[!t]
\centering
\begin{tikzpicture}
\definecolor{initcol}{RGB}{213,94,0}    
\definecolor{optcol}{RGB}{0,114,178}    
\definecolor{bandcol}{RGB}{0,114,178}   
\definecolor{infcol}{RGB}{230,159,0}    
\definecolor{threshcol}{RGB}{118,42,131}  

\begin{axis}[
    width=\columnwidth,
    height=0.8\columnwidth,
    xmode=log,
    xlabel={$\omega$ [rad/s]},
    ylabel={$\nu(\omega)$},
    xmin=1e0, xmax=1e5,
    ymin=-0.5, ymax=0.45,
    xtick={1e0, 1e1, 1e2, 1e3, 1e4, 1e5},
    ytick={-0.4,-0.2,0,0.2,0.4},
    xminorticks=false,
    yminorticks=false,
    axis lines=left,
    axis line style={-stealth, thick},
    tick align=outside,
    tick style={black, thick},
    label style={font=\small},
    tick label style={font=\small},
    grid=major,
    grid style={line width=0.1pt, draw=gray!12},
    major grid style={line width=0.2pt, draw=gray!18},
    enlargelimits=false,
    clip=true,
    axis on top=true,
    legend pos=north east,
    legend cell align=left,
    legend style={
        draw=gray!60,
        fill=white,
        fill opacity=0.92,
        text opacity=1,
        font=\scriptsize,
        inner sep=3pt,
        row sep=1pt,
        cells={anchor=west}
    }
]

\addplot[draw=none, fill=gray!7, forget plot]
coordinates {(35,-0.5) (35,0.45) (50,0.45) (50,-0.5)} \closedcycle;

\addplot[draw=none, fill=gray!7, forget plot]
coordinates {(35,-0.5) (50,-0.5) (50,0.0) (35,0.0)} \closedcycle;

\addplot[black, densely dashed, line width=0.8pt, forget plot]
coordinates {(1e0,0) (1e5,0)};

\addplot[black, densely dashed, line width=1.1pt, forget plot]
coordinates {(35,-0.5) (35,0.45)};
\addplot[black, densely dashed, line width=1.1pt, forget plot]
coordinates {(50,-0.5) (50,0.45)};

\addplot[initcol, densely dashed, line width=1.4pt]
table[x=omega, y=nu_0, col sep=comma]
{figures_data/mc_ifp.csv};
\addlegendentry{$\theta^{(0)}$}

\addplot[optcol, line width=1.7pt]
table[x=omega, y=nu_opt, col sep=comma]
{figures_data/mc_ifp.csv};
\addlegendentry{$\theta^{(13)}$}

\addplot[only marks, mark=*, mark size=2.0pt,
         mark options={fill=initcol, draw=white, line width=0.4pt}, forget plot]
coordinates {(35.4,0.2565)};
\addplot[only marks, mark=*, mark size=2.05pt,
         mark options={fill=optcol, draw=white, line width=0.4pt}, forget plot]
coordinates {(35.4,0.3184)};

\node[anchor=north, font=\scriptsize, text=gray!80!black]
    at (axis cs:550,-0.4) {$[\omega_1,\omega_2] { = } [35,50] \text{ [rad/s]}$};

\end{axis}
\end{tikzpicture}
\caption{Frequency-domain passivity profile of the microgrid
for the initial parameter set~$\theta^{(0)}$ (orange, dotted) and the
tuned set~$\theta^{(13)}$ (blue, solid), where
$\nu(\omega) = \lambda_{\min}\!\bigl((G(j\omega)+G^*(j\omega))/2\bigr)$.
The shaded gray region denotes the target frequency range~$\Omega = [35, 50]$\,rad/s.}
\label{fig:ifp_mc}
\end{figure}
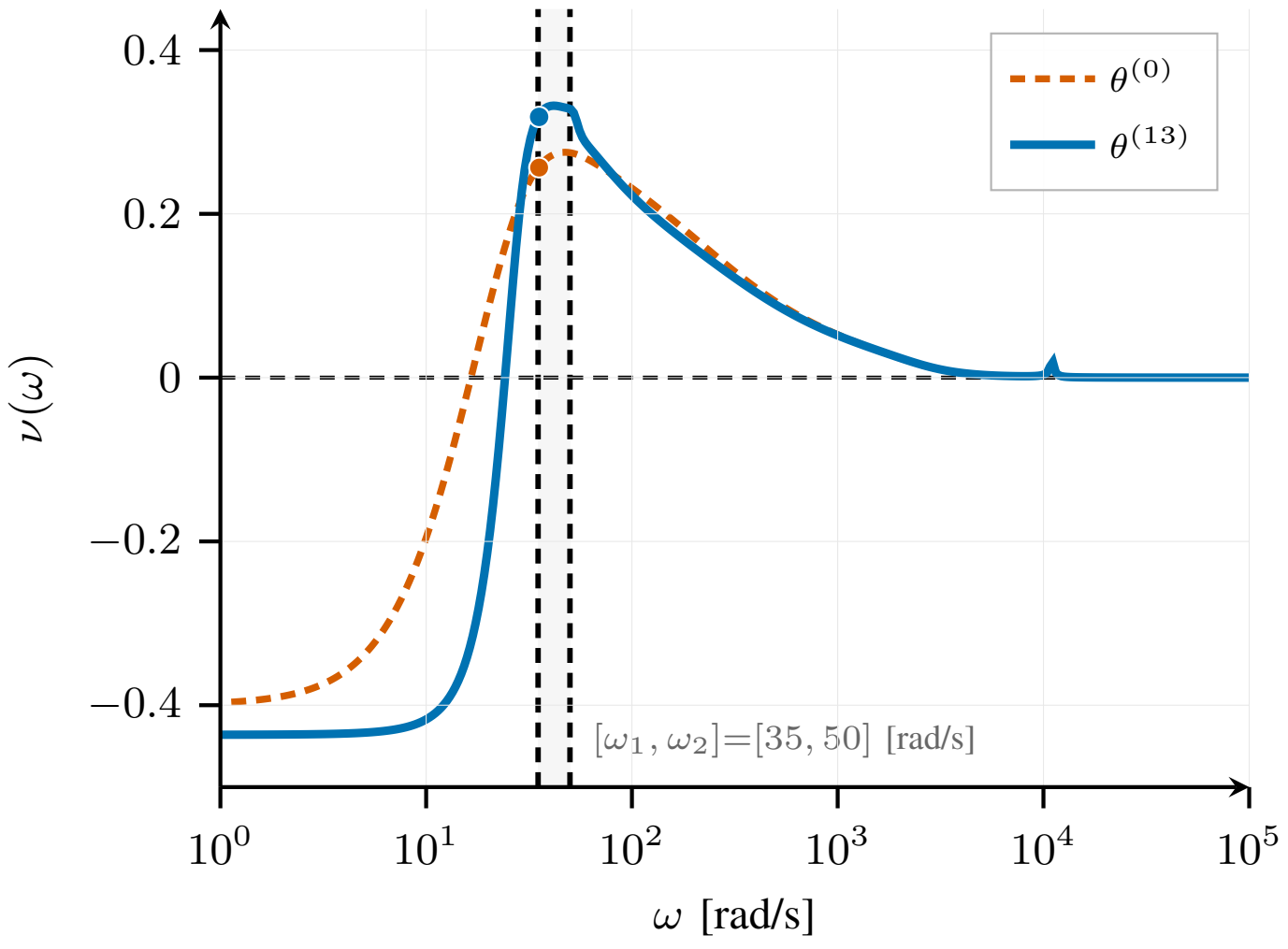
 
Figure~\ref{fig:ifp_mc} shows the resulting \ac{ifp} index over frequency. The
improvement is concentrated in the prescribed interval
$[35,50]$\,rad/s. Outside this interval, the margin may decrease. Indeed, below 35\,rad/s
the optimizer trades passivity for a larger index inside~$\Omega$. This trade-off confirms the frequency-selective nature of the proposed tuning procedure: the \ac{gkyp}-based index drives improvement within the targeted band, as the frequency-domain profile reflects.

\begin{figure}[!t]
\centering
\definecolor{initcol}{RGB}{213,94,0}   
\definecolor{optcol}{RGB}{0,114,178}   
\definecolor{rhpcol}{RGB}{213,94,0}    
\definecolor{trajcol}{RGB}{0,158,115}  
\definecolor{statcol}{RGB}{120,120,120}

\pgfplotsset{
    rl traj/.style={
        trajcol, line width=0.9pt, rounded corners=0.3pt,
        opacity=1, mark=*, mark size=0.9pt,
        mark options={fill=trajcol, draw=white, line width=0.2pt, solid},
    },
    theta0 marks/.style={
        color=initcol, only marks, mark=*, mark size=2.1pt,
        mark options={fill=initcol, draw=white, line width=0.4pt}
    },
    thetaopt marks/.style={
        color=optcol, only marks, mark=triangle*, mark size=2.7pt,
        mark options={fill=optcol, draw=white, line width=0.4pt}
    },
    stat marks/.style={
        color=statcol, only marks, mark=*, mark size=2.0pt,
        mark options={fill=statcol, draw=white, line width=0.3pt}
    },
    sep bar/.style={
        draw=none,
        legend image code/.code={%
            \draw[gray!60, line width=0.5pt] (0.2cm,-0.55ex) -- (0.2cm,1.4ex);%
        },
    },
    rl axis/.style={
        width=0.80\columnwidth,
        height=0.28\columnwidth,
        scale only axis=true,
        anchor=south west,
        axis lines=box,
        axis line style={black, semithick},
        xlabel={$\operatorname{Re}(\lambda)$~[rad/s]},
        ylabel={$\operatorname{Im}(\lambda)$~[rad/s]},
        xlabel style={font=\footnotesize, yshift=1pt},
        ylabel style={font=\footnotesize, yshift=-2pt},
        tick align=outside,
        tick style={black, semithick},
        tick label style={font=\scriptsize},
        grid=major,
        major grid style={line width=0.2pt, draw=gray!18},
        enlargelimits=false,
        clip=true,
        axis on top=false,
        title style={font=\footnotesize\bfseries, yshift=-2pt},
    }
}

\begin{tikzpicture}

\begin{axis}[
    rl axis,
    name=panelA,
    xlabel={},
    at={(0,0.82\columnwidth)},
    xmin=-2900, xmax=50,
    ymin=-13000, ymax=13000,
    xtick={-2500,-1500,-500},
    ytick={-10000,-5000,0,5000,10000},
    scaled y ticks=false,
    xticklabel style={/pgf/number format/fixed, /pgf/number format/set thousands separator={\,}},
    yticklabel style={/pgf/number format/fixed, /pgf/number format/set thousands separator={\,}},
    title={(a)~High-frequency modes},
    legend to name=mceiglegend,
    legend style={
        draw=gray!55,
        fill=white,
        font=\scriptsize,
        inner sep=3pt,
        column sep=5pt,
        cells={anchor=west},
    },
    legend columns=-1,
]

\draw[gray!45, line width=0.5pt] (axis cs:-2900,0) -- (axis cs:50,0);

\addplot[rl traj] table[x=real_1, y=imag_1, col sep=comma] {figures_data/mc_root_locus.csv};
\addlegendentry{path}
\addlegendimage{sep bar}
\addlegendentry{}
\addlegendimage{theta0 marks}
\addlegendentry{$\theta^{(0)}$}
\addlegendimage{sep bar}
\addlegendentry{}
\addlegendimage{thetaopt marks}
\addlegendentry{$\theta^{(13)}$}
\addlegendimage{sep bar}
\addlegendentry{}
\addlegendimage{stat marks}
\addlegendentry{$\approx$ unchanged}
\addplot[rl traj, forget plot] table[x=real_2, y=imag_2, col sep=comma] {figures_data/mc_root_locus.csv};
\addplot[rl traj, forget plot] table[x=real_3, y=imag_3, col sep=comma] {figures_data/mc_root_locus.csv};
\addplot[rl traj, forget plot] table[x=real_4, y=imag_4, col sep=comma] {figures_data/mc_root_locus.csv};
\addplot[rl traj, forget plot] table[x=real_5, y=imag_5, col sep=comma] {figures_data/mc_root_locus.csv};
\addplot[rl traj, forget plot] table[x=real_6, y=imag_6, col sep=comma] {figures_data/mc_root_locus.csv};
\addplot[rl traj, forget plot] table[x=real_7, y=imag_7, col sep=comma] {figures_data/mc_root_locus.csv};
\addplot[rl traj, forget plot] table[x=real_8, y=imag_8, col sep=comma] {figures_data/mc_root_locus.csv};
\addplot[rl traj, forget plot] table[x=real_9, y=imag_9, col sep=comma] {figures_data/mc_root_locus.csv};
\addplot[rl traj, forget plot] table[x=real_10, y=imag_10, col sep=comma] {figures_data/mc_root_locus.csv};
\addplot[rl traj, forget plot] table[x=real_11, y=imag_11, col sep=comma] {figures_data/mc_root_locus.csv};
\addplot[rl traj, forget plot] table[x=real_12, y=imag_12, col sep=comma] {figures_data/mc_root_locus.csv};
\addplot[rl traj, forget plot] table[x=real_13, y=imag_13, col sep=comma] {figures_data/mc_root_locus.csv};
\addplot[rl traj, forget plot] table[x=real_14, y=imag_14, col sep=comma] {figures_data/mc_root_locus.csv};
\addplot[rl traj, forget plot] table[x=real_15, y=imag_15, col sep=comma] {figures_data/mc_root_locus.csv};

\addplot[stat marks, forget plot,
    restrict expr to domain={\thisrow{real}}{-2900:-150}]
table[x=real, y=imag, col sep=comma] {figures_data/mc_eig_0.csv};


\end{axis}

\begin{axis}[
    rl axis,
    xlabel={},
    at={(0,0.41\columnwidth)},
    xmin=-145, xmax=2, ymin=-65, 
    ymax=65, 
    xtick={-120,-90,-60,-30,0}, 
    ytick={-60,-30,0,30,60},
    title={(b)~Low-frequency modes},
]

\draw[gray!45, line width=0.5pt] 
    (axis cs:-145,0) -- (axis cs:2,0);

\addplot[rl traj, forget plot] table[x=real_16, y=imag_16, col sep=comma] {figures_data/mc_root_locus.csv};
\addplot[rl traj, forget plot] table[x=real_17, y=imag_17, col sep=comma] {figures_data/mc_root_locus.csv};
\addplot[rl traj, forget plot] table[x=real_18, y=imag_18, col sep=comma] {figures_data/mc_root_locus.csv};
\addplot[rl traj, forget plot] table[x=real_19, y=imag_19, col sep=comma] {figures_data/mc_root_locus.csv};
\addplot[rl traj, forget plot] table[x=real_20, y expr=abs(\thisrow{imag_20}), col sep=comma] {figures_data/mc_root_locus.csv};
\addplot[rl traj, forget plot] table[x=real_21, y=imag_21, col sep=comma] {figures_data/mc_root_locus.csv};
\addplot[rl traj, forget plot] table[x=real_22, y=imag_22, col sep=comma] {figures_data/mc_root_locus.csv};
\addplot[rl traj, forget plot] table[x=real_23, y expr=-abs(\thisrow{imag_23}), col sep=comma] {figures_data/mc_root_locus.csv};

\addplot[theta0 marks,
    restrict expr to domain={\thisrow{real}}{-145:-2}]
table[x=real, y=imag, col sep=comma] {figures_data/mc_eig_0.csv};

\addplot[thetaopt marks,
    restrict expr to domain={\thisrow{real}}{-145:-2}]
table[x=real, y=imag, col sep=comma] {figures_data/mc_eig_opt.csv};

\end{axis}

\begin{axis}[
    rl axis,
    name=panelC,
    at={(0,0)},
    xmin=-3.5, xmax=1,
    ymin=-1, ymax=1,
    xtick={-3,-2,-1},
    extra x ticks={0},
    extra x tick labels={$0$},
    ytick={-1,0,1},
    title={(c)~Near-zero mode},
]

\addplot[draw=none, fill=rhpcol!10, forget plot]
coordinates {(0,-1) (1,-1) (1,1) (0,1)} \closedcycle;
\draw[rhpcol!70!black, densely dashed, line width=0.9pt]
    (axis cs:0,-1) -- (axis cs:0,1);
\node[rotate=90, anchor=center, font=\scriptsize\itshape, text=rhpcol!70!black]
    at (axis cs:0.5,0) {unstable (RHP)};

\draw[gray!45, line width=0.5pt] (axis cs:-3.5,0) -- (axis cs:1,0);

\addplot[rl traj, forget plot] table[x=real_24, y=imag_24, col sep=comma] {figures_data/mc_root_locus.csv};

\addplot[stat marks, forget plot,
    restrict expr to domain={\thisrow{real}}{-3.5:1}]
table[x=real, y=imag, col sep=comma] {figures_data/mc_eig_0.csv};


\end{axis}

\node[anchor=south, yshift=4pt]
    at (panelA.north |- current bounding box.north)
    {\pgfplotslegendfromname{mceiglegend}};

\end{tikzpicture}
\caption{Eigenvalues of the linearized microgrid
system matrix~$A$ for the initial parameter set~$\theta^{(0)}$ ($\bullet$, orange) and 
the tuned set~$\theta^{(13)}$ ($\blacktriangle$, blue).
Green dotted lines show the transition path across optimization iterations. 
Gray markers denote poles whose position remains essentially unchanged. 
All poles remain in the open left-half plane, confirming preserved stability.}
\label{fig:eig_mc}
\end{figure}
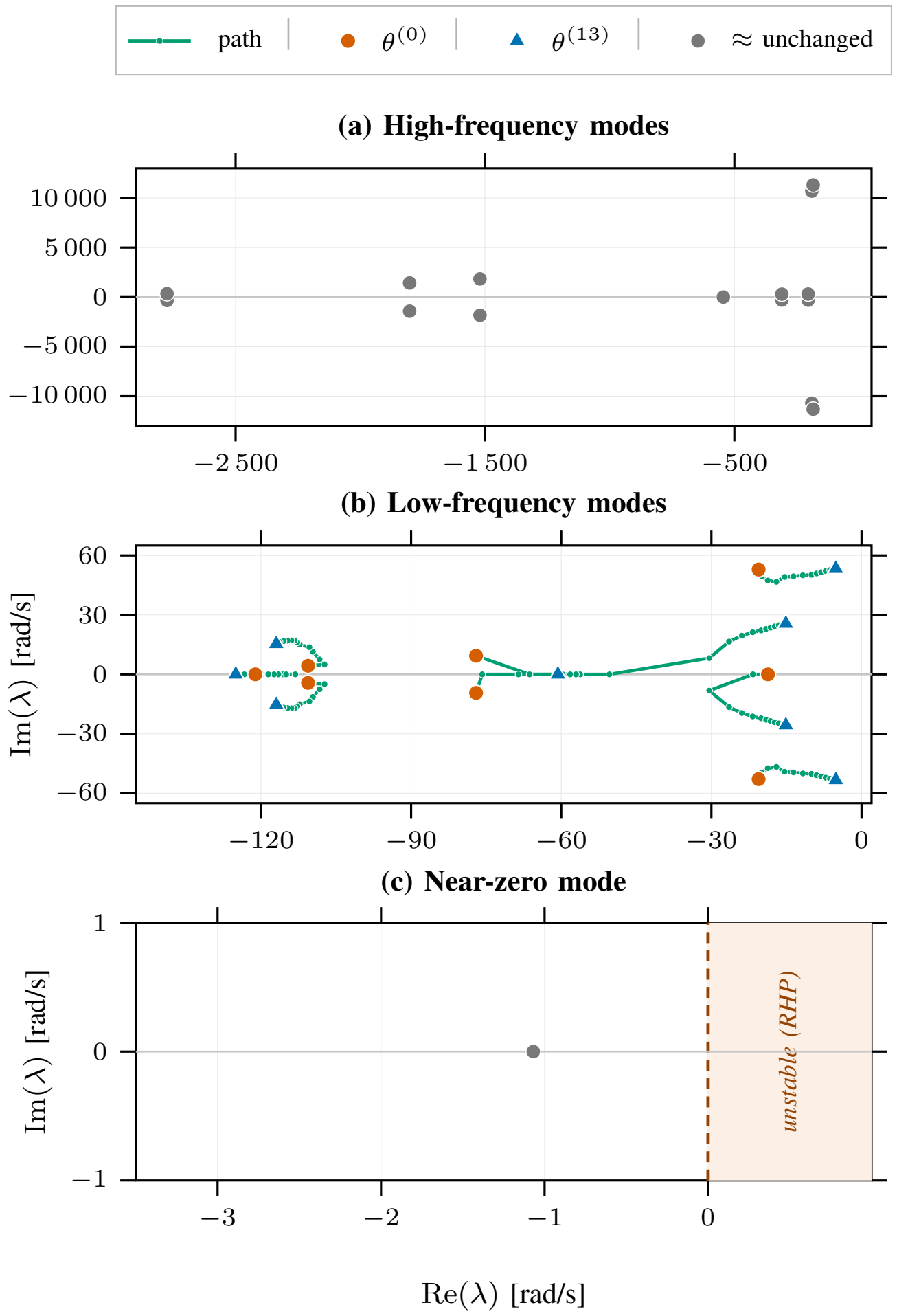
 
Figure~\ref{fig:eig_mc} shows the eigenvalue evolution over iterations. All 24 poles
remain in the open left-half plane during the optimization, consistent
with the Lyapunov stability constraint imposed in~\eqref{eq:stab_lmi}.
The most visible movements occur around the low-frequency modes ($f < 9$\,Hz),
consistent with the targeted frequency range for maximizing the \ac{ifp} index.
 
All in all, these examples confirm the effectiveness of the proposed method. The proposed method is able to automatically identify, using analytic derivatives via the envelope theorem, which tunable parameters to change for scenarios involving single assets as well as multiple interconnected assets with multiple tunable parameters.

\section{Conclusion}\label{sec:conclusion}

This paper presented a differentiable optimization framework for passivity-oriented controller tuning over prescribed frequency ranges. 
The proposed method combines a \ac{gkyp}-based optimization formulation with Envelope Theorem sensitivities,
enabling gradient-based tuning of frequency-dependent passivity indices.
The resulting bilevel algorithm tunes controller parameters while preserving Lyapunov stability at every iteration.

Numerical results on a grid-following \ac{vsc} and a multi-converter microgrid show that the method systematically improves the targeted passivity margin
over the selected frequency band, effectively shaping the passivity indices across the frequency domain. 
The framework therefore provides a structured alternative to trial-and-error tuning and parameter sweeps. 
The emerging regulatory trend toward frequency-band passivity requirements further underscores the practical relevance of the proposed method.

Future work will pursue two directions: experimental validation on hardware
converters and extension of the proposed framework to general \ac{gkyp}-based
control problems with formal convergence guarantees.

{\tiny
\setlength\bibitemsep{0pt}
\printbibliography
}



\appendix
\section{Mid-Frequency Sensitivity Derivation}

\section{SDPs with Complex Matrices} \label{app:sdps_complex}

The mid-frequency \ac{lmi} $M_{\text{MF}} \preceq 0$ is Hermitian 
because $\Psi_{\text{MF}}$ introduces imaginary off-diagonal entries 
(cf.\ Table~\ref{tab:psi_cases}). Since standard real-valued \ac{sdp} 
solvers require symmetric---not Hermitian---constraints, we first 
reformulate the problem in equivalent real form.

\begin{lemma}[Hermitian-to-real equivalence]\label{lem:complex_to_real}
    Let $X, Y \in \R^{n \times n}$ with $X = X^{\!\top}$ and 
    $Y = -Y^{\!\top}$, so that $M = X + jY \in \H^{n}$. Then
    \begin{equation*}
        X + jY \preceq 0
    \end{equation*}
    if and only if
    \begin{equation*}
        \begin{bmatrix} X & -Y \\ Y & X \end{bmatrix} \preceq 0.
    \end{equation*}
\end{lemma}

\begin{proof}
    By definition,
    \[
        X + jY \preceq 0 
        \;\iff\; 
        v^{*}(X + jY)\,v \leq 0, 
        \quad \forall\, v \in \mathbb{C}^n.
    \]
    Write $v = a + jb$ with $a, b \in \R^n$ and expand:
    \begin{align*}
        v^{*}(X + jY)\,v 
        &= (a^{\!\top} - jb^{\!\top})(X + jY)(a + jb) \\
        &= \underbrace{a^{\!\top}\!Xa + b^{\!\top}\!Xb 
           - a^{\!\top}\!Yb + b^{\!\top}\!Ya}_{\text{real part}} \\
        &\quad + j\underbrace{(a^{\!\top}\!Ya + a^{\!\top}\!Xb 
           - b^{\!\top}\!Xa + b^{\!\top}\!Yb)}_{\text{imaginary part}}.
    \end{align*}
    Symmetry of $X$ gives $a^{\!\top}\!Xb = b^{\!\top}\!Xa$; 
    skew-symmetry of $Y$ gives 
    $a^{\!\top}\!Ya = 0$ and $b^{\!\top}\!Yb = 0$. 
    Hence the imaginary part vanishes and
    \begin{align*}
        v^{*}(X + jY)\,v 
        &= a^{\!\top}\!Xa + b^{\!\top}\!Xb 
           - a^{\!\top}\!Yb + b^{\!\top}\!Ya \\
        &= \begin{bmatrix} a \\ b \end{bmatrix}^{\!\top}
           \begin{bmatrix} X & -Y \\ Y & X \end{bmatrix}
           \begin{bmatrix} a \\ b \end{bmatrix}.
    \end{align*}
    Setting $w = \begin{bmatrix} a \\ b \end{bmatrix} \in \R^{2n}$ 
    and noting that $w$ ranges over all of $\R^{2n}$ as $v$ ranges 
    over $\mathbb{C}^n$ completes the equivalence:
    \begin{align*}
        &v^{*}(X + jY)\,v \leq 0 \;\;\forall\, v \in \mathbb{C}^n \\
        &\quad \iff \quad
        w^{\!\top}
        \begin{bmatrix} X & -Y \\ Y & X \end{bmatrix} w \leq 0 
        \;\;\forall\, w \in \R^{2n} \\
        &\quad \iff \quad
        \begin{bmatrix} X & -Y \\ Y & X \end{bmatrix} 
        \preceq 0. \qedhere
    \end{align*}
\end{proof}

\subsection{Real-Valued Reformulation of the MF LMI}

We decompose the Hermitian constraint from 
Section~\ref{sec:freq_ranges} as 
$M_{\text{MF}} = X + jY$, where $X$ collects the real 
(symmetric) part and $Y$ the imaginary (skew-symmetric) part. 
The block entries read
\begin{align*}
X_{11} &= A^{\!\top}\!P + PA - A^{\!\top}\!QA 
- \bar{\omega}_1\bar{\omega}_2\, Q + \Sigma_{11}, \\
X_{12} &= PB - A^{\!\top}\!QB + \Sigma_{12}, \\
X_{22} &= -B^{\!\top}\!QB + \Sigma_{22},
\end{align*}
with $X_{21} = X_{12}^{\!\top}$, and
\begin{align*}
Y_{11} &= \bar{\omega}_c\bigl(A^{\!\top}\!Q - QA\bigr), \\
Y_{12} &= -\bar{\omega}_c\, QB, \\
Y_{22} &= 0,
\end{align*}
with $Y_{21} = -Y_{12}^{\!\top}$. 
Applying Lemma~\ref{lem:complex_to_real}, the equivalent real 
constraint is
\begin{equation}\label{eq:mf_real_lmi}
M_{\text{MF}}^{\mathbb{R}} :=
\begin{bmatrix} X & -Y \\ Y & X \end{bmatrix} \preceq 0,
\end{equation}
a real symmetric matrix of dimension $2(n+m) \times 2(n+m)$.

\subsection{MF Sensitivity via the Envelope Theorem} \label{app:mf_sensitivity}

We derive the sensitivity of $\alpha^{\star}(\theta)$ for 
Problem~\eqref{eq:inner_sdp} with the real-valued MF 
constraint~\eqref{eq:mf_real_lmi} replacing the LF/HF \ac{lmi}.
Let $Z_M^{\mathbb{R}} \succeq 0$ denote the dual multiplier 
associated with $M_{\text{MF}}^{\mathbb{R}} \preceq 0$. 
Theorem~\ref{thm:envelope_sdp} yields
\begin{equation}\label{eq:envelope_mf}
\frac{\partial \alpha^{\star}}{\partial \theta_k}\bigg|_{\bar\theta}
=
-\left\langle
Z_M^{\mathbb{R}\star}(\bar\theta),\;
\frac{\partial M_{\text{MF}}^{\mathbb{R}}}{\partial \theta_k}
\bigg|_{\bar\theta}
\right\rangle,
\quad k=1,\dots,p.
\end{equation}
It remains to derive the analytical formula of \(\frac{\partial M_{\text{MF}}^{\mathbb{R}}}{\partial \theta_k}
\bigg|_{\bar\theta}\).
The block structure of~\eqref{eq:mf_real_lmi} is preserved under 
differentiation:
\begin{equation}\label{eq:dMmf_structure}
\frac{\partial M_{\text{MF}}^{\mathbb{R}}}{\partial \theta_k}
\bigg|_{\bar\theta}
=
\begin{bmatrix}
\partial_k X & -\partial_k Y \\[4pt]
\partial_k Y & \partial_k X
\end{bmatrix},
\end{equation}
where $\partial_k X := \partial X/\partial\theta_k|_{\bar\theta}$ 
and $\partial_k Y := \partial Y/\partial\theta_k|_{\bar\theta}$. 
We compute each in turn, applying the chain rule and evaluating at 
$\bar\theta$.

\subsubsection{Derivative of the Real Part}

\begin{align*}
(\partial_k X)_{11}
&= (\partial_k A)^{\!\top} P^{\star} + P^{\star}\,(\partial_k A)
- \bigl[(\partial_k A)^{\!\top} Q^{\star} A 
+ A^{\!\top} Q^{\star}\,(\partial_k A)\bigr]
\\
&\quad + \rho^{\star}\bigl[(\partial_k C)^{\!\top} C 
+ C^{\!\top}\,(\partial_k C)\bigr],
\\[6pt]
(\partial_k X)_{12}
&= P^{\star}\,(\partial_k B)
- \bigl[(\partial_k A)^{\!\top} Q^{\star} B 
+ A^{\!\top} Q^{\star}\,(\partial_k B)\bigr]
\\
&\quad - \tfrac{1}{2}(\partial_k C)^{\!\top}
+ \rho^{\star}\bigl[(\partial_k C)^{\!\top} D 
+ C^{\!\top}\,(\partial_k D)\bigr],
\\[6pt]
(\partial_k X)_{22}
&= -\bigl[(\partial_k B)^{\!\top} Q^{\star} B 
+ B^{\!\top} Q^{\star}\,(\partial_k B)\bigr]
\\
&\quad - \tfrac{1}{2}\bigl(\partial_k D + (\partial_k D)^{\!\top}\bigr)
+ \rho^{\star}\bigl[(\partial_k D)^{\!\top} D 
+ D^{\!\top}\,(\partial_k D)\bigr],
\end{align*}
with $(\partial_k X)_{21} = (\partial_k X)_{12}^{\!\top}$.

\subsubsection{Derivative of the Imaginary Part}

\begin{align*}
(\partial_k Y)_{11}
&= \bar{\omega}_c\bigl[(\partial_k A)^{\!\top} Q^{\star} 
- Q^{\star}\,(\partial_k A)\bigr],
\\[6pt]
(\partial_k Y)_{12}
&= -\bar{\omega}_c\, Q^{\star}\,(\partial_k B),
\\[6pt]
(\partial_k Y)_{22}
&= 0,
\end{align*}
with $(\partial_k Y)_{21} = -(\partial_k Y)_{12}^{\!\top}$.
\\
Substituting~\eqref{eq:dMmf_structure} into~\eqref{eq:envelope_mf} 
and expanding the trace yields the mid-frequency passivity index sensitivity. 
Collecting all $p$ components forms the gradient 
$\nabla_\theta\alpha^{\star}(\bar\theta)$.

\end{document}